\documentclass[prd,superscriptaddress,preprintnumbers,amsmath,amssymb,nofootinbib]{revtex4-1}

\usepackage{amsmath}
\usepackage{amssymb}
\usepackage{slashed}
\usepackage{graphicx}
\usepackage{graphics}
\usepackage{epsfig}
\usepackage{color}
\usepackage{xcolor}
\usepackage{soul}
\usepackage{hyperref}
\usepackage[section]{placeins}
\usepackage{mathrsfs}
\usepackage{dsfont}
\usepackage[normalem]{ulem}

\newcommand{\be}{\begin{equation}}
\newcommand{\ee}{\end{equation}}
\newcommand{\bea}{\begin{eqnarray}}
\newcommand{\eea}{\end{eqnarray}}

\newcommand{\dcrit}{\mathbf{d}_{\mathbf{\mathrm c}}}

\newcommand{\cA}{\mathcal A}
\newcommand{\cN}{\mathcal N}
\newcommand{\bd}{\mathbf{d}}
\newcommand{\bc}{\mathbf{c}}
\newcommand{\vol}{\mathcal V}

\begin{document}
\title{Spectral dimension on spatial hypersurfaces in causal set quantum gravity}

\author{Astrid Eichhorn}
   \email{eichhorn@sdu.dk}
   \affiliation{CP3-Origins, University of Southern Denmark, Campusvej 55, DK-5230 Odense M, Denmark}
\affiliation{Institut f\"ur Theoretische
  Physik, Universit\"at Heidelberg, Philosophenweg 16, 69120
  Heidelberg, Germany}

\author{Fleur Versteegen}
  \email{f.versteegen@thphys.uni-heidelberg.de}
\affiliation{Institut f\"ur Theoretische
  Physik, Universit\"at Heidelberg, Philosophenweg 16, 69120
  Heidelberg, Germany}
  
  \author{Sumati Surya}
  \affiliation{Raman Research Institute, C.V. Raman Avenue, Sadashivanagar, Bangalore
560 080, India}

\begin{abstract} 
    An important probe of quantum geometry is its spectral dimension, defined via
      a spatial diffusion process. In this work we study the spectral dimension of a ``spatial hypersurface''  in a manifoldlike  causal set using the induced spatial distance function. In  previous work, the diffusion was taken on the full causal set, where the nearest neighbours are unbounded in number.  The resulting super-diffusion leads to an increase in the spectral dimension at short diffusion times, in contrast to other approaches to quantum gravity. In the current work, by using a temporal  localisation in  the causal set,  the number of nearest {\it spatial} neighbours is rendered finite. Using numerical simulations of causal sets obtained from $d=3$ Minkowski spacetime, we find that for a  flat spatial hypersurface,   the spectral dimension agrees with the Hausdorff dimension at intermediate scales,  but shows clear indications of dimensional reduction at small scales, i.e., in the ultraviolet. The latter is a direct consequence of ``discrete asymptotic silence''  at small scales in causal sets.
\end{abstract}

\maketitle

\section{Introduction}
In any theory of quantum gravity it is  important to understand the properties of quantum geometry. Due to quantum fluctuations of spacetime, properties like the dimension can differ from their classical counterparts. 
The spectral dimension is a well-explored dimension indicator in quantum-gravity models \cite{Ambjorn:2005db,Laiho:2016nlp,Benedetti:2008gu,Benedetti:2009ge,Anderson:2011bj,Lauscher:2005qz,Reuter:2011ah,Rechenberger:2012pm,Calcagni:2013vsa,Horava:2009if,Calcagni:2014cza,Eichhorn:2013ova}, see \cite{Carlip:2009kf,Carlip:2017eud,Carlip:2019onx} for overviews. It is obtained by considering the diffusion of a ficiticious test particle on the geometry, with the dimension being extracted from the return probability.
 In the case of a flat, classical background, the (non-relativistic) diffusion equation in a $d$-dimensional space $X$  is given by
\be
\left(\partial_{\sigma}- \partial_x^2 \right)P(x,x',\sigma)=0,\label{eq:diffeq}
\ee
where $x',x$ are the initial and final positions of the test particle in $X$, $\sigma$ is an external diffusion time \footnote{This is different from the proper time, or coordinate time of the test particle in the spacetime and should not be confused with these.} and $P(x,x',\sigma)$ denotes  the probability distribution for the test particle to go from   $x'$ to $x$ in diffusion time $\sigma$. Here the diffusion constant has been absorbed using an appropriate choice of units, \footnote{For a more general space of course, the diffusion depends on the location of the test particle. Here, we are looking at the simplest case, so that there is a position independent diffusion constant, and the diffusion equation is  just the heat equation.
} so that the solution can be expressed as 
\be
P(x,x',\sigma) = \frac{1}{(4\pi\sigma)^{d/2}}e^{-\frac{(x-x')^2}{4\sigma}}.\label{eq:probd}
\ee
The return probability is then  $P_r(\sigma)=  \frac{1}{V}\int_V P(x,x,\sigma) =(4\pi\sigma)^{-\frac{d}{2}}$  which is the probability for the  fictitious test particle to return to the starting point after diffusion time $\sigma$. Thus, in flat spacetime one has the scaling relation $P_r \propto \sigma^{-\frac{d}{2}}$.  This suggests the more general definition for  the  spectral dimension $d_s$ as the scaling dimension of $P_r(\sigma)$
  \begin{equation} 
    P_r(\sigma) \approx \sigma^{-\frac{d_s}{2}}\quad \Rightarrow \quad 
    d_s(\sigma) = - 2\frac{d \ln P_r(\sigma)}{d\ln \sigma},\label{eq:dscont}
  \end{equation}
  even when $P_r(\sigma)$ is not of the simple form above. Thus, in flat spacetime $d_s$ is the same as the Hausdorff dimension, and is independent of the diffusion time $\sigma$.

On the other hand for quantum geometries $d_s(\sigma)$ can depend on the scale at which the effective spatial geometry is probed. For large enough scales it should give the Hausdorff dimension, while at small  scales (the ultraviolet regime) it can show a significant deviation.
Multiple quantum-gravity approaches
exhibit a dimensional reduction in $d_s$ in the ultraviolet regime, e.g., \cite{Benedetti:2008gu,Ambjorn:2005db,Laiho:2016nlp,Benedetti:2009ge,Anderson:2011bj,Lauscher:2005qz,Reuter:2011ah,Rechenberger:2012pm,Calcagni:2013vsa,Horava:2009if,Calcagni:2014cza}.

Causal set quantum-gravity \cite{Bombelli:1987aa} (see \cite{Sorkin:2003bx,Dowker:2005tz,Henson:2010aq,Dowker:aza,Surya:2019ndm,Wallden:2013kka} for reviews) provides an intriguing exception: explicit simulations of the diffusion process on causal sets which are sprinklings into both Minkowski and de Sitter spacetimes
instead show a dimensional {\sl increase}  at small $\sigma$  \cite{Eichhorn:2013ova}.
In that setup, the diffusion process proceeds along the links (both future- and past- directed), i.e., the irreducible causal relations between elements in a causal set.  Unlike a regular lattice, however continuum-like  causal sets, which are directed acyclic graphs, are not of fixed or even   finite valency.  For example for a causal set that is approximated by Minkowski spacetime, the valency is infinite as a consequence of Lorentz invariance, which can be viewed as a form of non-locality intrinsic to causal sets. This leads to superdiffusion even in the presence of an infrared cut-off:  given a large number of nearest neighbours, a diffusing particle has a very large number of possibilities to quickly diffuse away from its starting point,  resulting in a sharp decrease of the return probability with diffusion time and, consequently, a large spectral dimension at small diffusion times. The same is true for a causal diffusion process that only proceeds along links in a future-directed fashion  where the dimension can be extracted from the meeting probability of two diffusion processes  \cite{Eichhorn:2013ova}. 
This dimensional increase has been associated with asymptotic silence \cite{Carlip:2015mra}. As shown in \cite{Eichhorn:2017djq},   manifoldlike causal sets do indeed exhibit a discrete asymptotic silence regime, but  this in turn can be used to define a new dimension estimator, which does show dimensional reduction. The key point is then to be able to identify accurately the correct discrete spectral dimension estimator for a causal set. We note here that 
the  causal-set d'Alembertian \cite{Benincasa:2010ac,Benincasa:2010as,Dowker:2013vba,Aslanbeigi:2014zva} can also be used to ``upgrade" the classical diffusion equation Eq.~\eqref{eq:diffeq}. In  \cite{Carlip:2015mra,Belenchia:2015aia} the \emph{continuum} approximation of the causal set d'Alembertian was used to analytically estimate the spectral dimension at length scales smaller than an intermediate ``non-locality'' scale. In that regime the spectral dimension exhibits dimensional reduction. However, for causal sets, when the non-locality scale is chosen to be the same as the discreteness scale, the corresponding diffusion equation no longer has an interpretation, and one must instead look at the diffusion process on the causal set.  For the nonlocality scale much larger than the discreteness scale, the microscopic process underlying the diffusion equation in \cite{Carlip:2015mra,Belenchia:2015aia}  is a different one from the random walk implemented in  \cite{Eichhorn:2013ova}.

The dimensional increase seen in the work of \cite{Eichhorn:2013ova} is deeply linked to Lorentz invariance, and hence one might imagine that the appropriate diffusion equation must be one that is Lorentz invariant \footnote{In \cite{Dowker:2003hb,Philpott:2008vd}, a Lorentz invariant diffusion process was used to obtain  a phenomenological model of particles diffusing in momentum space. }.  Whether such a process can give rise to a new Lorentzian spectral dimension is an interesting open question. Instead, in this present work, we find the appropriate setting for the non-relativistic diffusion process in a causal set. Rather than the full causal set, we consider the analogue of a spatial hypersurface in the causal set, which is an {\sl inextendible antichain, $\mathcal{A}$}. We use  a class of induced spatial distance functions  on $\mathcal{A}$, obtained in \cite{Eichhorn:2018doy} to define a finite set of nearest neighbours for the diffusion process, thus removing the possibility of super-diffusion. This non-relativistic diffusion might be interpreted as the limiting case of a physical process by making a rough split of the  causal set into ``moments of time'' $\mathcal{A}_t$. Considering a diffusion process on space instead of spacetime brings the analysis closer to that in other quantum-gravity approaches, where one either explores diffusion on a discrete spatial configuration, or directly studies a spatial diffusion equation.

\section{Spatial diffusion on a causal set}
Causal set quantum gravity is based on the insight that the information encoded in a causal spacetime metric on a given manifold is equivalent to the causal structure plus a local volume element. Accordingly, each causal spacetime is a partially ordered set of spacetime points, as the causal relation $\prec$ (``precedes'') defines a partial order. For any spacetime without closed timelike curves, the order relation satisfies the following properties for all elements $x,\, y,\, z$ in the spacetime 

\begin{equation*}
\begin{aligned}
 &\mbox{ i)  Transitivity: } x\prec y \mbox{ and } y\prec z \Rightarrow x \prec z,\\
 &\mbox{ ii)  Acyclicity: } x\prec y \mbox{ and } y\prec x \Rightarrow x =y.
 \end{aligned}
 \end{equation*}
 
 In causal set quantum gravity, the additional assumption of spatiotemporal discreteness of each causal set $C$ is made, implying a locally finite cardinality: ${\rm card}\{z \in C\, |\, x<z<y \}< \infty$.  The continuum plays no fundamental role in causal sets in the sense that it is viewed as an emergent, approximate description of spacetime at a sufficiently coarse scale. To relate the phenomenological consequences of causal set quantum gravity to known physics which is typically based on a continuum description, it is useful to be able to relate a given causal set to a continuum manifold and vice-versa.
 Given a continuum spacetime \footnote{Henceforth spacetime is assumed to be causal.}, a corresponding causal set can be constructed via a process called Poisson sprinkling. This  is a random selection of causal-set elements from the spacetime, following the Poisson probability. This choice for the probability distribution leads to a number-volume correspondence on average, with the fluctuations kept to a minimum \cite{Saravani:2014gza}. The causal relations in the causal set are inherited from the continuum metric. The converse construction of a continuum manifold corresponding to a causal set (at sufficiently coarse-grained scales) is an in general unsolved problem. A continuum manifold is said to approximate a given causal set if that causal set has a high probability to emerge from the manifold via the sprinkling process.  Several methods to check for manifold-likeness have been developed for two-dimensional causal sets in e.g., \cite{Henson:2006dk,Aghili:2018tae} as well as for arbitrary dimension \cite{Bolognesi:2014vpa,Glaser:2013pca,Major:2009cw}. Furthermore it was shown in  \cite{Bombelli:2006nm} that a causal set obtained through a sprinkling into Minkowski spacetime remains Lorentz invariant in the continuum approximation.
 
In causal set quantum gravity, the quantisation proceeds via the path-integral formalism. A candidate dynamics, the Benincasa-Dowker action \cite{Benincasa:2010ac,Benincasa:2010as,Dowker:2013vba} provides a starting point to evaluate the path integral via Monte Carlo simulations  \cite{Surya:2011du,Glaser:2014dwa,Glaser:2017sbe}. In this paper, we will not consider the path integral over all causal sets. Instead, we will evaluate the spectral dimension on sprinklings into three-dimensional Minkowski spacetime. Thus, ours is a purely kinematical study in the sense that we neglect the effect of the causal-set dynamics and only average over sprinklings into $\mathbb{M}^3$ with a constant and equal weight.

\begin{figure}
\centering
\includegraphics[width=0.7\linewidth,clip=true, trim=0cm 11cm 7cm 7cm]{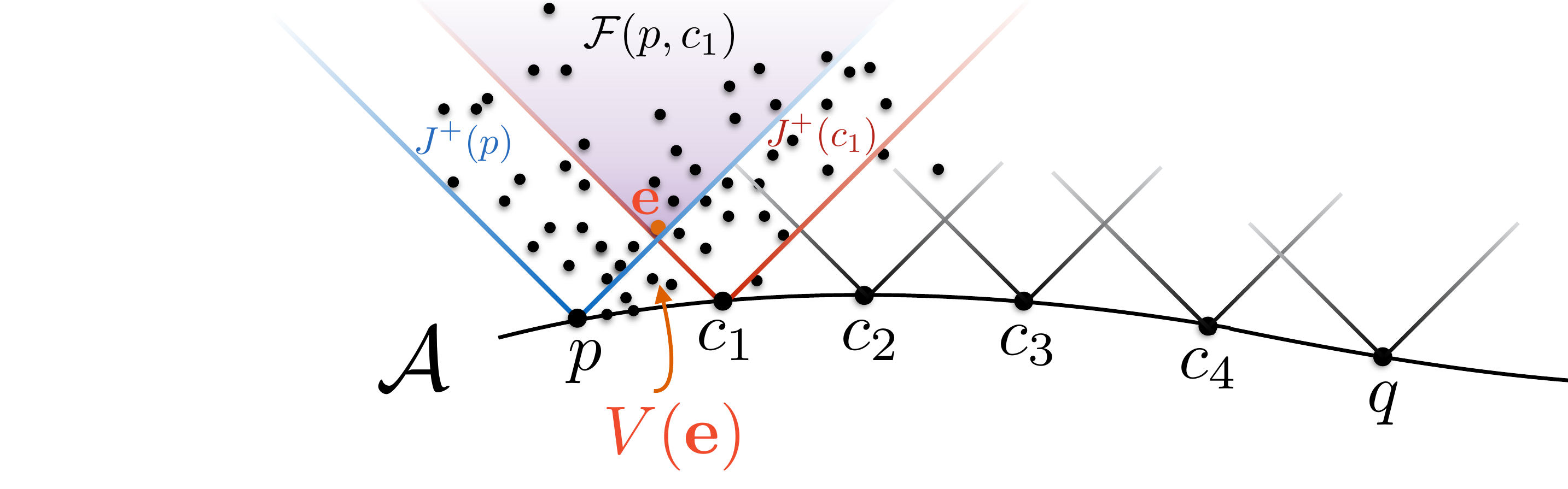}
\caption{\label{fig:illustration_distance}We illustrate the construction of the distance between $p$ and $q$ using the causal volume to the future of the antichain $\mathcal{A}$.}
\end{figure}

To set up a spatial diffusion process, we do not use the full causal set, in which Lorentz invariance dictates that there are an infinite number of nearest neighbours, as was done in \cite{Eichhorn:2013ova}. Instead, we consider a ``spatial hypersurface'' in the causal set  and use a spatial-distance function defined in \cite{Eichhorn:2018doy} to define the nearest neighbours for the diffusion process. In a causal set, the analogue of a spatial hypersurface is a collection of (causally) unrelated spacetime points, called an antichain $\mathcal{A}$. The spatial distance on such an -- a priori structureless -- antichain can be evaluated using causal information, as shown in \cite{Eichhorn:2018doy}.
In spirit, this idea is related to the reconstruction of spatial topology from a thickening of the antichain, i.e., the inclusion of the past and/or future of the antichain, as in \cite{Major:2006hv,Major:2009cw}.
In essence, the key idea is  as follows: Given two unrelated elements $p$, $q$ in a causal set $C$, the intersection of their causal futures, $\mathcal{F}(p,q)=J^{+}(p){\bigcap} J^+(q)$ forms a subset of $C$. For every element $r \in \mathcal{F}(p,q)$, one can calculate the intersection of the causal past of $r$ with the causal future of the antichain $\mathcal{A}$, $J^-(r)\bigcap J^+(\mathcal{A})$, where $J^+(\mathcal{A}) =\bigcup_{i\in \mathcal{A}}J^+(i)$, cf.~Fig.~\ref{fig:illustration_distance}.
Next, one minimises the  volume of this intersection, i.e., ${\rm Vol}(J^-(r)\bigcap J^+(\mathcal{A}))$ over all $r$, to find 
\be
V({\mathbf{e}})\equiv \underset{r\in \mathcal{F}(p,q)}{\rm inf} {\rm Vol}(J^-(r)\bigcap J^+(\mathcal{A})).
\ee
 For $n$-dimensional Minkowski spacetime, $V({\mathbf{e}})$ is directly related to the induced spatial distance between $p$ and $q$ for a flat (inertial) spatial hypersurface
 \be
 \widetilde{\mathbf{d}}(p,q) = 2 \left( \frac{V({\mathbf{e}})}{\zeta_n}\right)^{\frac{1}{n}}.\label{eq:predistance}
 \ee
 For more general spatial hypersurfaces this relation receives curvature corrections which are small when $p$ and $q$ are ``close'' as shown in \cite{Eichhorn:2018doy}.
$\widetilde{\mathbf{d}}(p,q)$ provides a ``predistance function'' on $\cA$ which then must be minimised. Recognising the importance of curvature effects, this flat spacetime formula can however only make sense for small enough regions of the embedding spacetime. Hence, one introduces a meso-scale cut-off,  $V(\mathbf{e}) \leq v_m$, so that $\widetilde{\mathbf{d}}(p,q)$ is defined only when $p,q$ are closer than $d_m$, where we use the flat spacetime relation  $v_m = \zeta_{n} d_m^{n}$, and $\zeta_{n}= \frac{\pi^{(n-1)/2}}{n\Gamma\left( \frac{n+1}{2}\right)}$. This gives a predistance function on $\cA$. To obtain the full distance function on $\cA$, $\widetilde{\mathbf{d}}(p,q)$  minimised over all ``paths'' in $\cA$ from $p$ to $q$.  A $k$-element path $\chi_k(p,q)$ from $p$ to $q$ is defined as the sequence  of elements $ (c_0,c_1,....,c_{k-1},c_k)$, $c_0=p, c_k=q$, $c_i \in \cA$ ,  such that  $\widetilde{\mathbf{d}}(c_i,c_{i+1}) < d_m$ for all $i \in (0,1, \ldots, k-1)$. The possible values of $k$ depend both on the choice of $p,q \in \cA$ as well as the  mesoscale $d_m$.  Defining the path distance  as (cf.~Fig.~\ref{fig:illustration_distance}) 
\be
\widetilde{\mathbf{d}}_{\chi_k}(p,q)\equiv\sum_{i=0}^{k-1}\tilde{\mathbf d}(c_i,c_{i+1}), 
\ee
the spatial distance function is  obtained by minimising over the set $\Gamma(p,q)$ of  all possible paths  in $\cA$ between $p$ and $q$ 
\be
\mathbf{d}(p,q) \equiv \underset{\chi \in \Gamma(p,q)}{\rm inf} \widetilde{\mathbf{d}}_{\chi}.
\ee
As shown in \cite{Eichhorn:2018doy},  if the local curvature scale and the discreteness scale are sufficiently separated, there is a large range of meso-scales for which this distance function is ``stable'' or independent of the cut-off, and also corresponds to the continuum induced distance at scales larger than the discreteness scale. Around the discreteness scale it deviates significantly from the continuum, exhibiting  discrete asymptotic silence \cite{Eichhorn:2017djq}.

This spatial distance can be used to define the nearest neighbours of an element on $\cA$ and hence set up a diffusion process on $\cA$. The distance function  renders $\cA$ into a complete graph, but with weighted edges. If one were to define the nearest neighbours of an element  $e \in \cA$ as the set of ``closest'' elements, then this would likely give just a single neighbour, which would give rise to an almost deterministic process. We also know that at the smallest distances, the effect of the discreteness scale is felt via asymptotic silence. Thus a sensible definition would be to choose the nearest neighbours of $e$ to be those that lie within a certain radius from $e$,
  \begin{equation}
\cN_\bc(e) \equiv \{ e' \in \cA| \bd(e,e') \leq \dcrit \}. 
\end{equation}
Clearly, because of the inherent randomness of a continuum-like causal set, the cardinality of the set $\cN_c(e)$ varies from element to element on $\cA$ even for fixed $\dcrit$. Thus, even if the valency of our resultant graph is finite, it is not fixed. 
In \cite{Eichhorn:2018doy} a correlation between the number of nearest neighbours $|\cN_c(e)|$ and the continuum induced spatial volume of the ball of radius $\dcrit$ was used to obtain the Hausdorff dimension $d_H$ of the spatial hypersurface approximating $\cA$ for $\dcrit$ larger than the scale of asymptotic silence. Specifically,  the number of nearest neighbours $|\cN_c(e)|$ was found to scale with the distance as $\sim\dcrit^{d_H}$.
\begin{figure}[!t]
\includegraphics[width=0.4\linewidth]{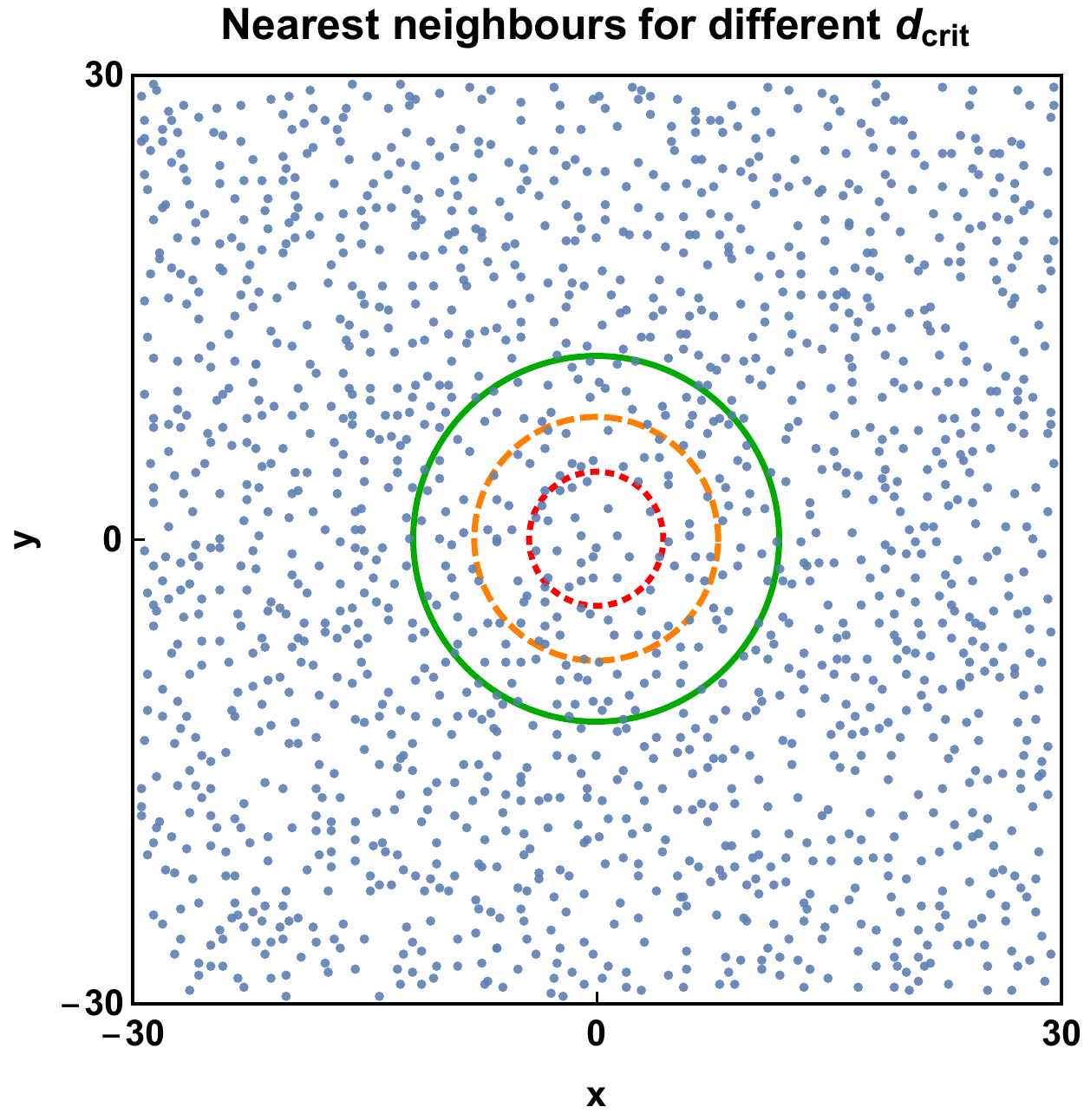}
\caption{\label{fig:nearneigh_illu} Illustration of the number of nearest neighbours for an element in the centre of the antichain (blue dots). The red dotted circle identifies all the nearest neighbours for $\dcrit \approx 4.3$, the orange dashed contains the nearest neighbours for $\dcrit \approx 7.9$ and the green continuous circle for $\dcrit \approx 11.8$, respectively. Note that the larger values of $\dcrit$ include the smaller regions, i.e., for $\dcrit \approx 11.8$ all the elements within all three circles are accessible from the origin in one step. }
\end{figure}

The diffusion process on $\cA$ is set up with respect to an external diffusion time $\sigma$, which is not necessarily related to the time intrinsic to the causal set.  At each integer diffusion time, the  fictitious test particle takes a step to one of the nearest neighbours of the current element with uniform probability. As the valency is not constant, the probability depends on the element $e$, and is given by $1/n(e)$, where $n(e)$ is the number of nearest neighbours of $e$. For our simulations we include $e$ itself as one of its nearest neighbours.  The choice of uniform distribution is for practical reasons. Given the distance function it might seem odd to weigh every nearest neighbour equally, given that they lie within a solid ball centred on $e$. However, because of the numbers involved, the simulations are largely insensitive to this weight. From a geometric point of view what we have done is to take the weighted complete graph $\cA$ and replace it with an effective graph $G(\dcrit)$ of finite, but not fixed valency which depends on the choice of $\dcrit$. The diffusion process proceeds  according to the graph distance on $G(\dcrit)$.  We provide examples in Sec.~\ref{sec:results}.

\begin{figure}[!t]
\includegraphics[width=0.4\linewidth]{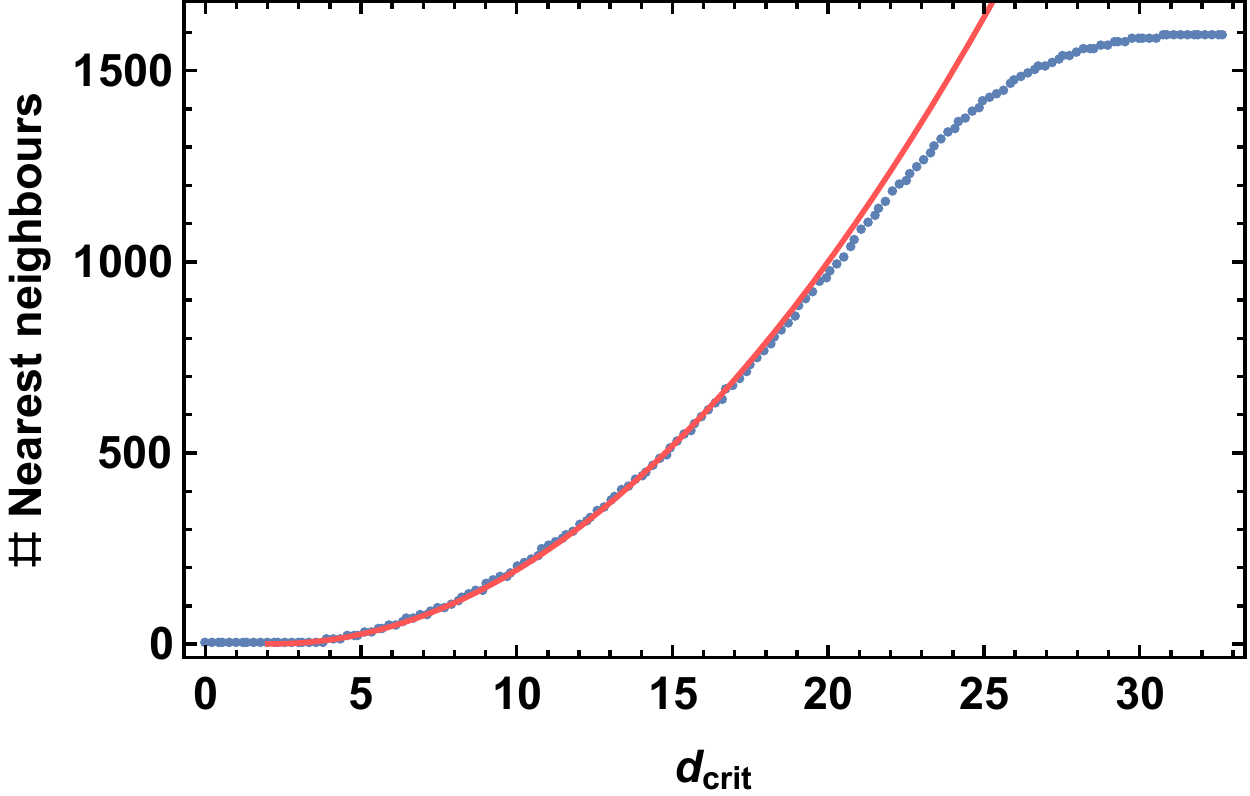}
\caption{\label{fig:nnnscaling}We show the number of nearest neighbours within distance $\dcrit$ to an element $e$ in the center of the antichain (blue dots). The red line shows the expected scaling according to the Hausdorff dimension, $n = 1+ \pi(r-r_0)^2$. Herein, $r_0=2.166$ is the critical distance above which an element has more than one nearest neighbour. (Each element is also one of its own nearest neighbours.)}
\end{figure}

The diffusion process depends on the choice of  $\dcrit$ which is a free parameter. Varying over $\dcrit$ one can explore various regimes, from the ultraviolet (UV), small-scale limit to the infrared (IR), large-scale limit. For  $\dcrit$ much larger than the discreteness scale, one expects continuum behaviour to emerge. In particular, in the continuum approximation, one expects the spectral dimension $d_s$ to approach the Hausdorff dimension $d_H$ of the spatial hypersurface. Defining the  volume $\vol_e(d_c)$ of a ball of radius $d_c$ on $G(\dcrit)$  to be  the number of elements within a graph distance $d_c$ of an element $e$, in the continuum limit one expects the scaling  $\vol_e(d_c) \propto d_c^{d_H}$, which corresponds to usual definition of the Hausdorff dimension in the continuum.
As shown in Fig.~\ref{fig:nnnscaling} the number of nearest neighbours of $e$ grows with $\dcrit^2$ for an antichain approximated by a two-dimensional spatial hypersurface, up to a value of $\dcrit$ where boundary effects set in due to the finite size of the sprinkling. Accordingly, the continuum dimension $d_H$ is encoded in the connectivity (number of nearest neighbours), and therefore the spectral dimension is expected to approach the Hausdorff dimension for large enough $\dcrit$.

Conversely, the regime of small $\dcrit$ is expected to exhibit dimensional reduction. This is intuitively clear in the limit $\dcrit \rightarrow 0$, where no element of the antichain has a nearest neighbour other than itself. Therefore, the effective graph $G(\dcrit)$  is completely disconnected. Diffusion cannot take place on such a disconnected set. Hence, the return probability in this limit is constant, as all random walkers are confined to their respective starting points, so that 
\be
d_s=0,
\ee
 which is  an extreme form of dimensional reduction. This is directly related to discrete asymptotic silence, discussed in \cite{Eichhorn:2017djq}, where there is an effective  ``decoupling'' of worldlines in the UV. Specifically, the discrete spatial distance overestimates the continuum distance in this case, since the first common element to the future  of any  two elements in $\cA$ lies further in their future in the causal set compared to the continuum. Accordingly, for $\mathbf{d}$  of the order of the discreteness scale (more precisely, about an order of magnitude larger, see \cite{Eichhorn:2017djq}), $\cA$ contains significantly fewer elements at spatial distance $\mathbf{d}$ than in  the continuum. The drastic reduction in the  number  of nearest neighbours in the  UV  leads to an effective  decoupling of worldlines, and also manifests itself in a vanishing spectral dimension.

We expect that as
 $\dcrit$ increases, small  connected ``islands" will be formed in the antichain, which confine the diffusion process. From Fig.~\ref{fig:nnnscaling}, obtained from sprinklings into $\mathbb{M}^3$, we see that even for small spatial distances, the scaling of the number of nearest neighbours with $\dcrit$ reproduces the Hausdorff dimension. However, the local connectivity of each ``island" is necessary but not sufficient to extract a stable value of the spectral dimension. In addition to exhibiting the appropriate local connectivity, the spatial hypersurface used in the diffusion process also needs to be sufficiently large to delay the onset of boundary effects to  large diffusion times. Since for very small values of $\dcrit$ asymptotic silence confines the diffusion to a small subset of the original antichain, boundary effects set in for small $\sigma$ when the diffusion process runs into the boundary of its island. The fact  that each diffusion process can only reach a small subset of elements (irrespective of the number of steps it can take), implies that at small $\dcrit$, it cannot recover features of an \emph{unbounded} continuum spacetime. 
 Thus, we expect that  the spectral dimension will underestimate the continuum dimension in this regime.
The boundary effects induced by asymptotic silence decrease with increasing $\dcrit$, such that the spectral dimension is expected to transition from 0 to 2, as we increase $\dcrit$ from 0 on sprinklings into $\mathbb{M}^3$, where the continuum dimension of spatial hypersurfaces is 2.

\section{Implementation}
We consider ten different sprinklings into 2+1 dimensional Minkowski spacetime $\mathbb{M}^3$, with antichains approximating a flat spatial hypersurface, with an average of 1600 points. We calculate the spatial distance between all pairs of points according to the prescription discussed above. 
On each antichain, we pick 30 different starting points and start up to $7\cdot 10^4$ diffusion processes (less for smaller values of $\dcrit$). Using a  uniform random number generator, the diffusion process  randomly selects one of its nearest neighbours in each diffusion step.
To check that our  {\verb!C++!} code works as expected, and to obtain the requisite number of diffusion processes for the results to converge, we use the exact probability as a benchmark.

\begin{figure}
\includegraphics[width=0.6\linewidth]{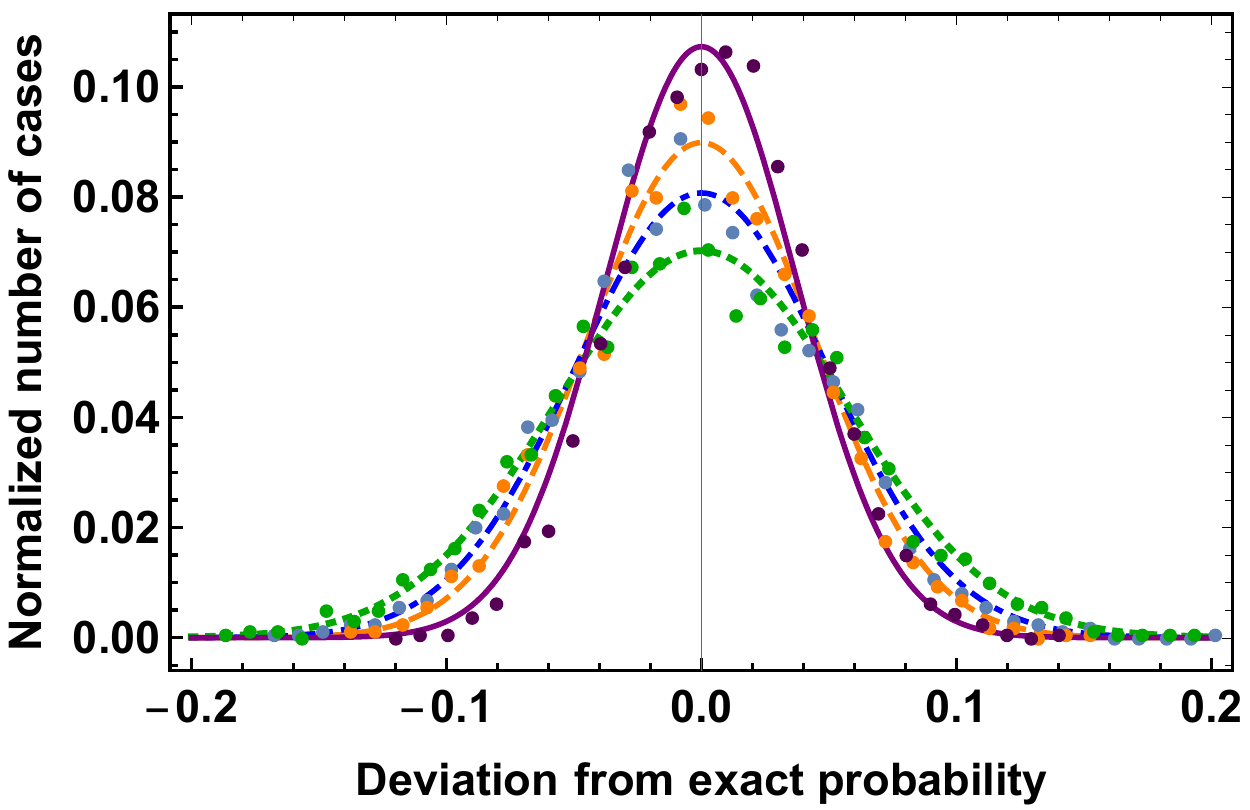}
\caption{\label{fig:stats} We show the normalised number of cases for a given deviation from the exact probability for $3\cdot 10^4$ diffusion processes (green dots), $4\cdot 10^4$ (blue dots), $5\cdot 10^4$ (orange dots) and $7\cdot 10^4$ (purple dots). We also show Gaussian fits to the data (green dotted for $3\cdot 10^4$, blue dashed-dotted for $4\cdot 10^4$ orange dashed for $5\cdot 10^4$ and purple continuous for $7\cdot 10^4$). As expected for a properly implemented diffusion algorithm, the Gaussian becomes more narrow and more strongly peaked, as the number of diffusion processes is increased.}
\end{figure}
Specifically, given an element $e$, we compare the exact probability to jump to any one of its nearest neighbours, which is $1/n(e)$, to the fraction of cases in which the diffusion process jumps to that particular nearest neighbour. As expected, for a small  number of diffusion processes, the fraction of cases in which the diffusion process jumps to a particular nearest neighbour deviates from $1/n(e)$ due to statistical fluctuations, while for a large enough number of diffusion processes, the deviation between the two decreases, cf.~Fig.~\ref{fig:stats}. We use this to find the requisite number of diffusion processes, as shown in Tab.~\ref{tab:diffusers}. 
\begin{table}
\begin{tabular}{c|c|c|c|c|c|c|c|c}
$\dcrit$ & 3.5 & 3.9 & 4.3 & 5.9 & 6.9 & 7.9 &9.8 & 11.8\\\hline
number of diffusion processes & $3\cdot 10^4$& $3\cdot 10^4$&$3\cdot 10^4$& $5\cdot 10^4$& $5\cdot 10^4$& $7\cdot 10^4$&$7\cdot 10^4$&$7\cdot 10^4$\\\hline
\end{tabular}
\caption{\label{tab:diffusers}The number of diffusion processes used for each value of $\dcrit$ for a given starting element and sprinkling. For each of the 10 sprinklings used, the diffusion process was started from 30 different elements in the antichain $\cA$. The total number of diffusion processes for a given $\dcrit$ is thus given by $300$ times the number quoted in the table.}
\end{table}
For the  range of values of   $\dcrit$ shown in Tab.~\ref{tab:diffusers} the number of nearest neighbours transitions from $n(e)=1$ to $n(e)\gg 1$, cf.~Fig.~\ref{fig:nnnscaling}, making this the most interesting regime to study the $\dcrit$-dependence of $d_s$.

To arrive at the final return probability, we average over all diffusion processes, starting points and sprinklings. 
Here we restrict ourselves to causal sets obtained by sprinklings into $\mathbb{M}^3$. However, the diffusion process can be run on an inextendible antichain in any causal set, which would be required to perform the full path integral in causal-set quantum gravity.
As computation time scales quadratically with the linear extent of the spatial hypersurface, exploring very large hypersurfaces is rather expensive. Therefore, we limit ourselves to hypersurfaces with about 1600 elements. This limits the values of  $\dcrit$ since there is an onset of boundary effects already at low diffusion times $\sigma$ for larger values of $\dcrit$.

\begin{figure}[!t]
\includegraphics[width=0.4\linewidth]{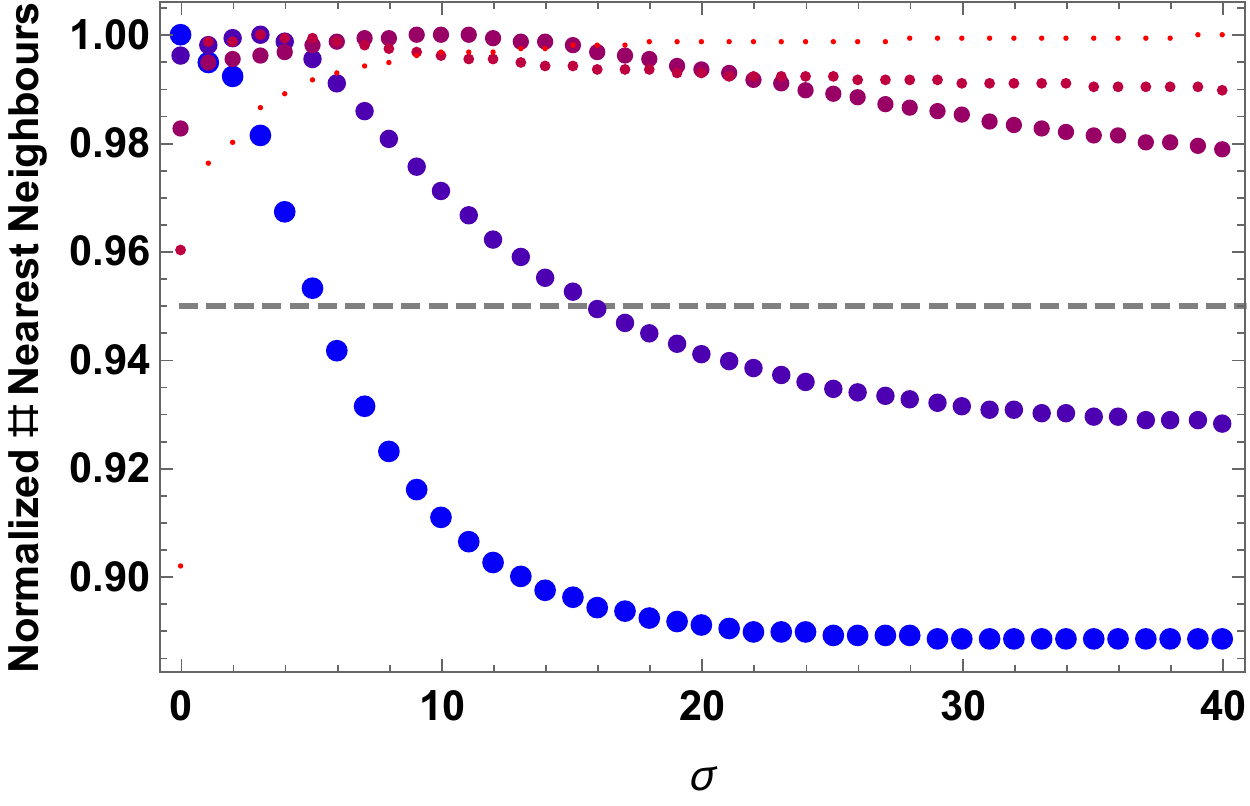}
\caption{\label{fig:boundary} Normalised number of nearest neighbours as a function of the diffusion time $\sigma$ for various values of $\dcrit$. The increase in $\dcrit$ is indicated by an increase in the size of the circles and a change of colour from red to blue. The smallest red circles correspond to $\dcrit \approx 3.5$ followed by $\dcrit \approx 4.3,\, 5.9,\, 7.9,\, 9.8$. The gray, dashed line at $0.95$ helps determine  for which value of $\sigma$ a loss of $5\%$  in the  average number of nearest neighbours has occurred. When this line is crossed, boundary effects become important. For $\dcrit \approx 3.5,\, 4.3,\, 5.9 $ no boundary effects are expected up to $\sigma = 40$, whereas for $\dcrit \approx 7.9,\, 9.8$ boundary effects already set in at $\sigma \approx 17 $ and $\sigma \approx 7$, respectively. We thus stop the diffusion processes at these corresponding values of $\sigma$.}
\end{figure}

To make sure that our results are not contaminated by boundary effects, we identify the onset of boundary effects as follows.
Elements near the boundary have on average fewer  nearest neighbours than elements in the centre of a causal set. Thus, a drop in the average number of nearest neighbours can be used to identify the diffusion time $\sigma_{bd}(\dcrit)$ at which boundary effects set in for a given $\dcrit$. We impose the conservative condition that a drop of $\sim5\%$ in the number of nearest neighbours signals the onset of boundary effects and stop the diffusion processes at the corresponding value of $\sigma$. For larger $\dcrit$, this limits the reliable range of diffusion times quite severely, as described in   Fig.~\ref{fig:boundary}.

\section{Results}\label{sec:results}

\begin{figure}
\includegraphics[width=0.45\linewidth]{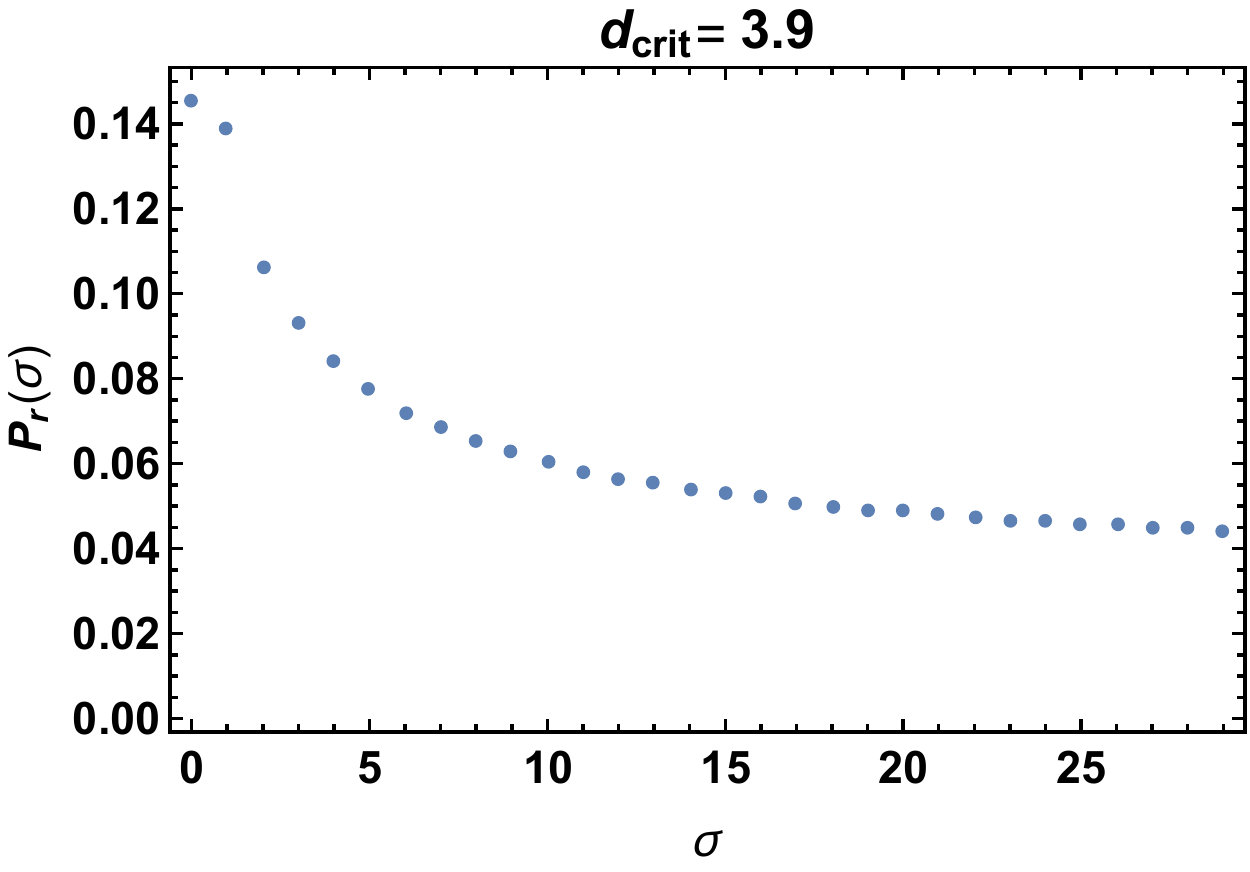}\quad\includegraphics[width=0.45\linewidth]{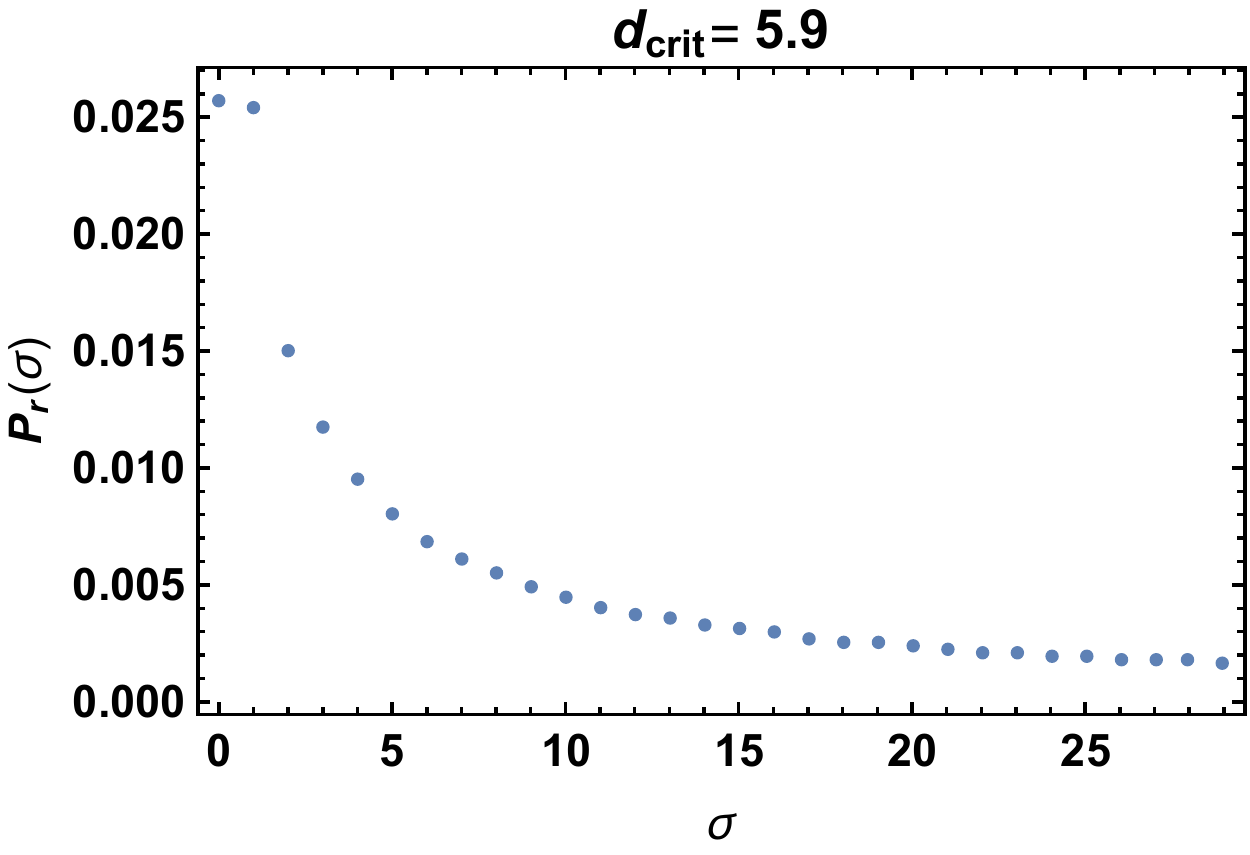}\newline
\includegraphics[width=0.45\linewidth]{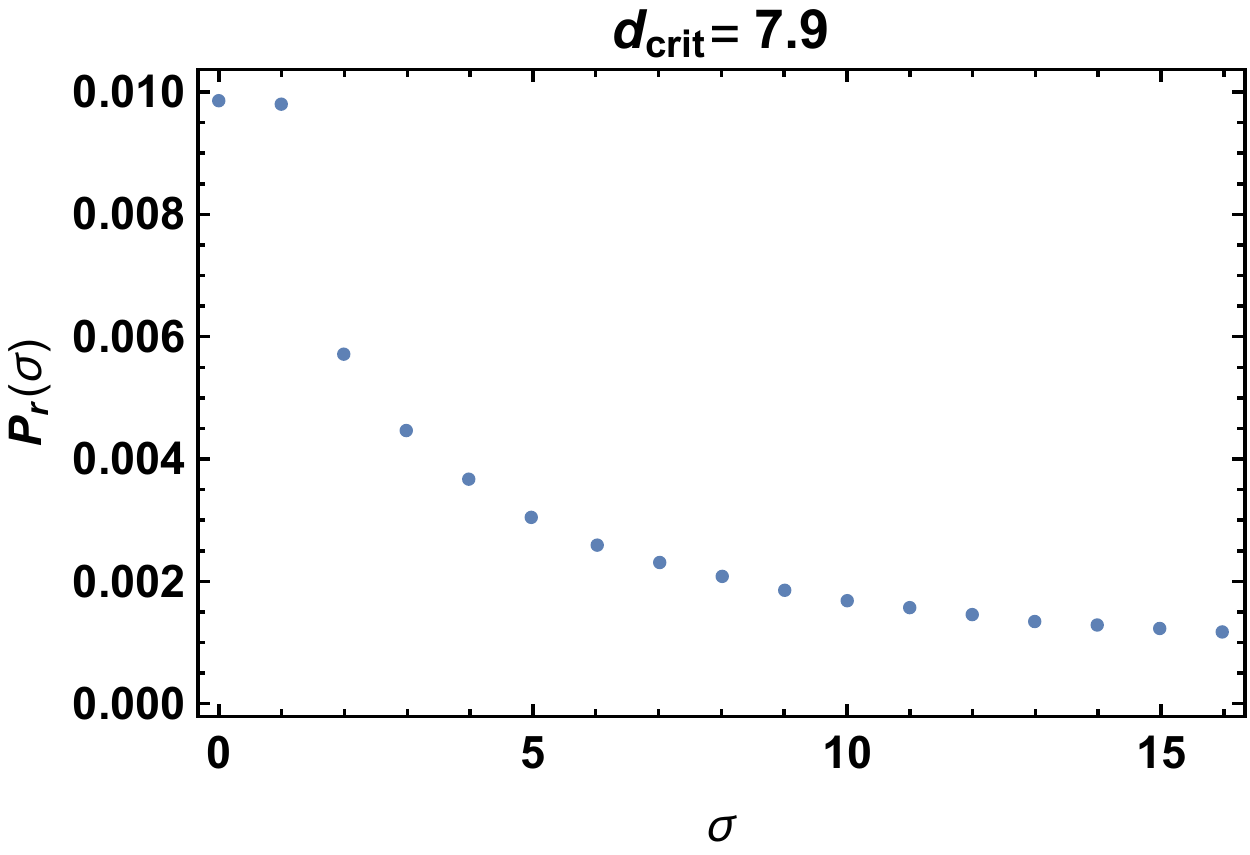}\quad\includegraphics[width=0.45\linewidth]{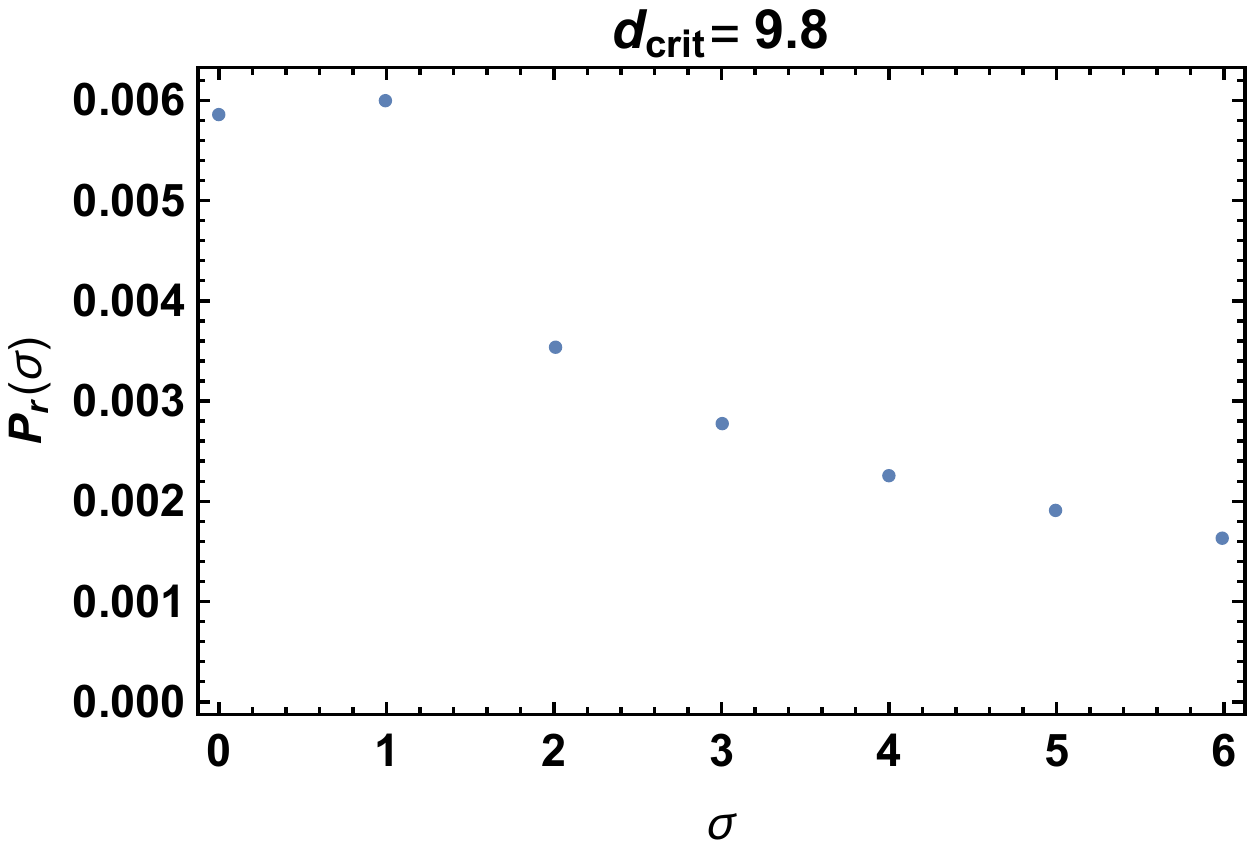}
\caption{\label{fig:Pr_new}  We show the numerical data for the return probability for the cases $\dcrit = 3.9$ (left upper panel), $\dcrit=5.9$ (right upper panel), $\dcrit=7.9$ (left lower panel) and $\dcrit = 9.8$ (right lower panel).} 
\end{figure}

We obtain numerical estimates for the return probability averaged over starting points and sprinklings at the discrete time steps $\sigma=0,1,2,...$  as explained above. As shown in Fig.~\ref{fig:Pr_new},
 the return probability remains constant or changes only very slightly in the first diffusion step. This is a consequence of the fact that in the 0th time step, the probability for the diffusion process to remain at the origin, $o$, is $1/n(o)$. In the second step, the total return probability is the sum of the probabilities to jump back from any of the neighbours of $o$. On average, each neighbour has $n(o)$ neighbours itself. Accordingly, the return probability remains essentially unchanged in the very first step. Beyond $\sigma=1$, the probability to return to the origin decreases rapidly at early times. This results in a finite value for the spectral dimension, which is related to minus the slope of the return probability. At  larger diffusion times $\sigma$, the return probability tends towards a constant for all values of $\dcrit$, resulting in a drop-off of $d_s$ to zero.  This is a consequence of the finite extent of the antichain, which results in an equilibration of the diffusion process in a finite amount of time. The latter behaviour is a consequence of boundary effects. As the size of the causal set is increased, one expects the associated drop in $d_s$ to occur at a larger value of $\sigma$.
 
\begin{figure}[t!]
\includegraphics[width=0.32\linewidth]{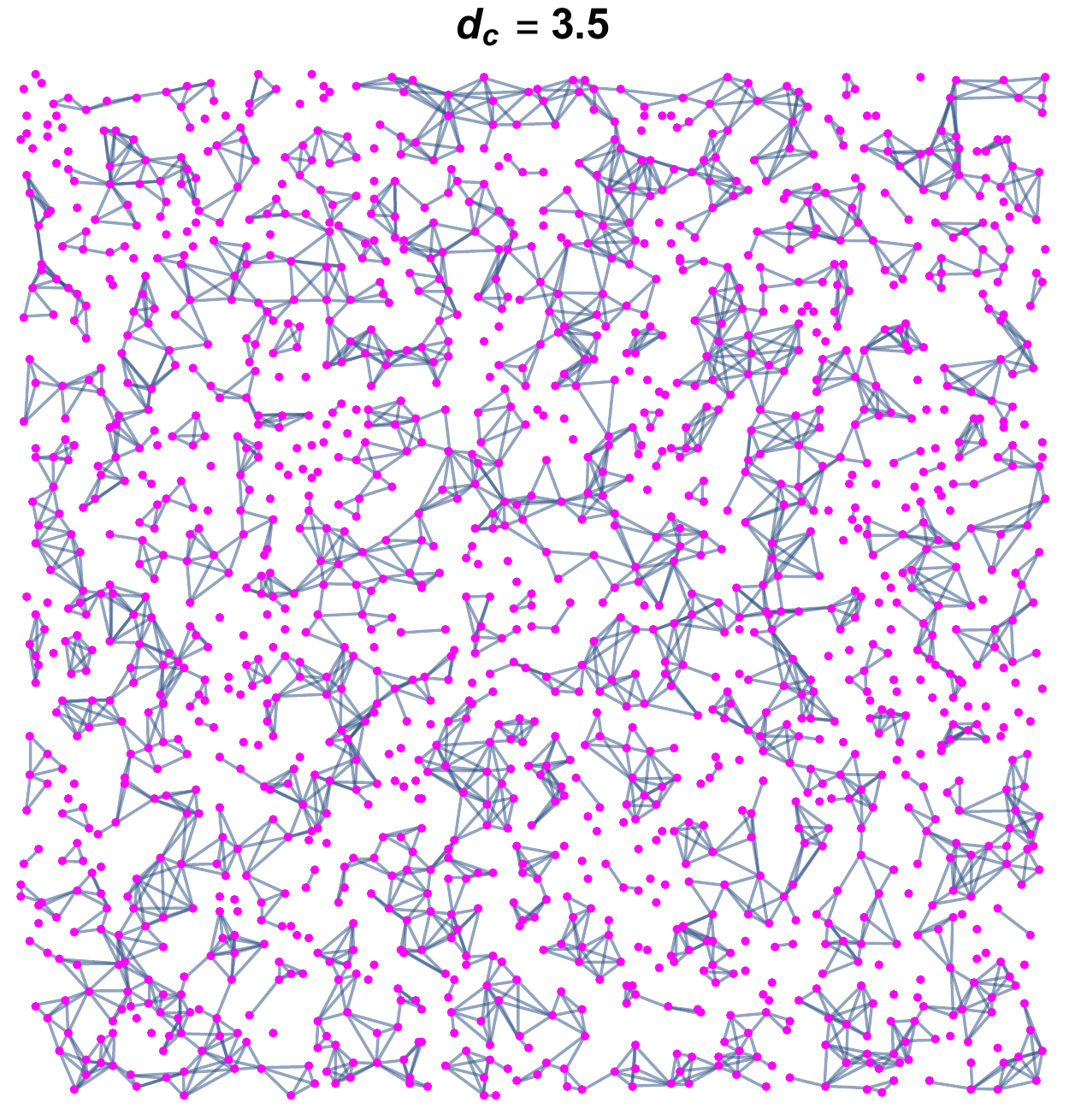}\quad
\includegraphics[width=0.32\linewidth]{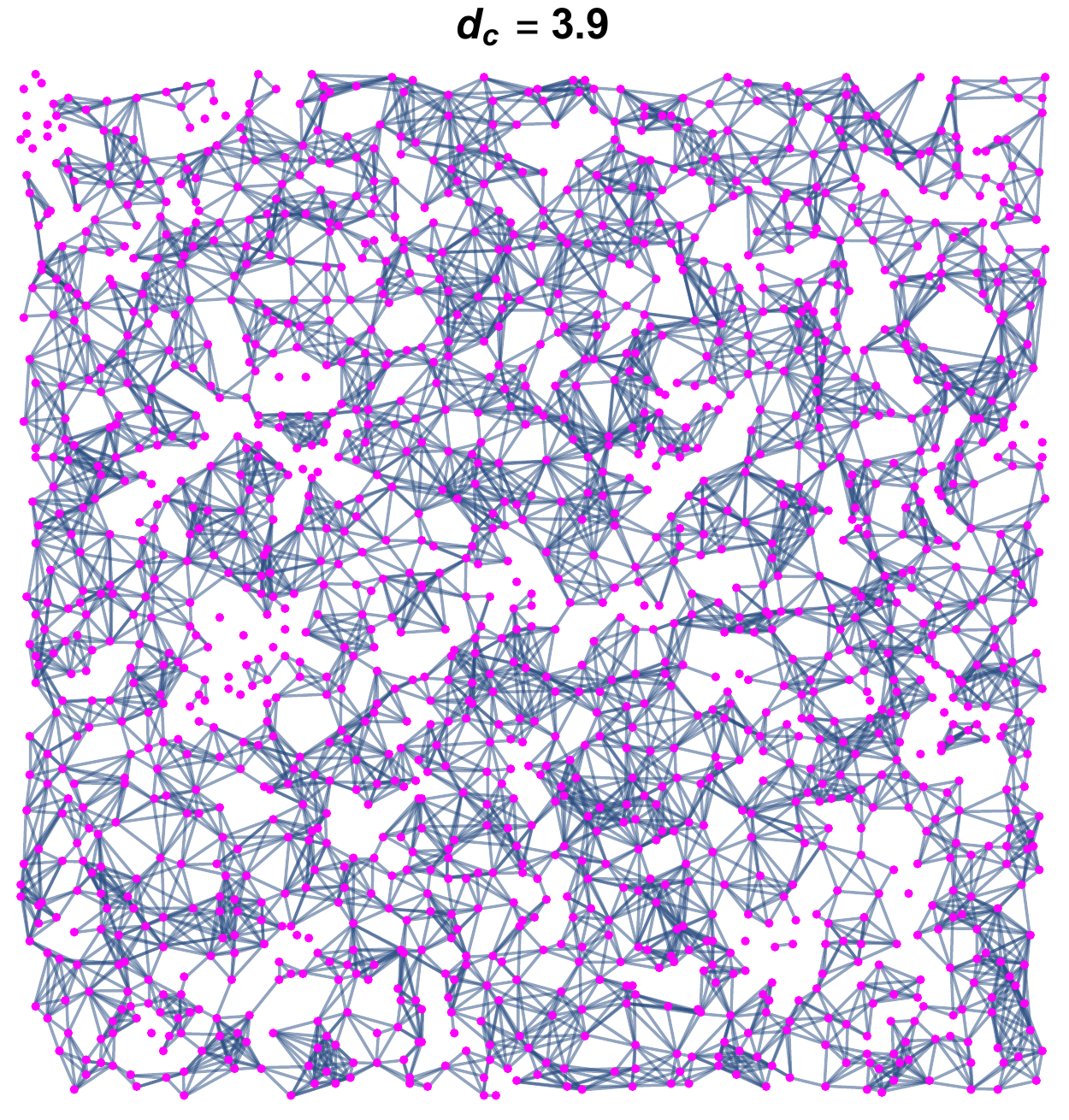}\quad
\includegraphics[width=0.32\linewidth]{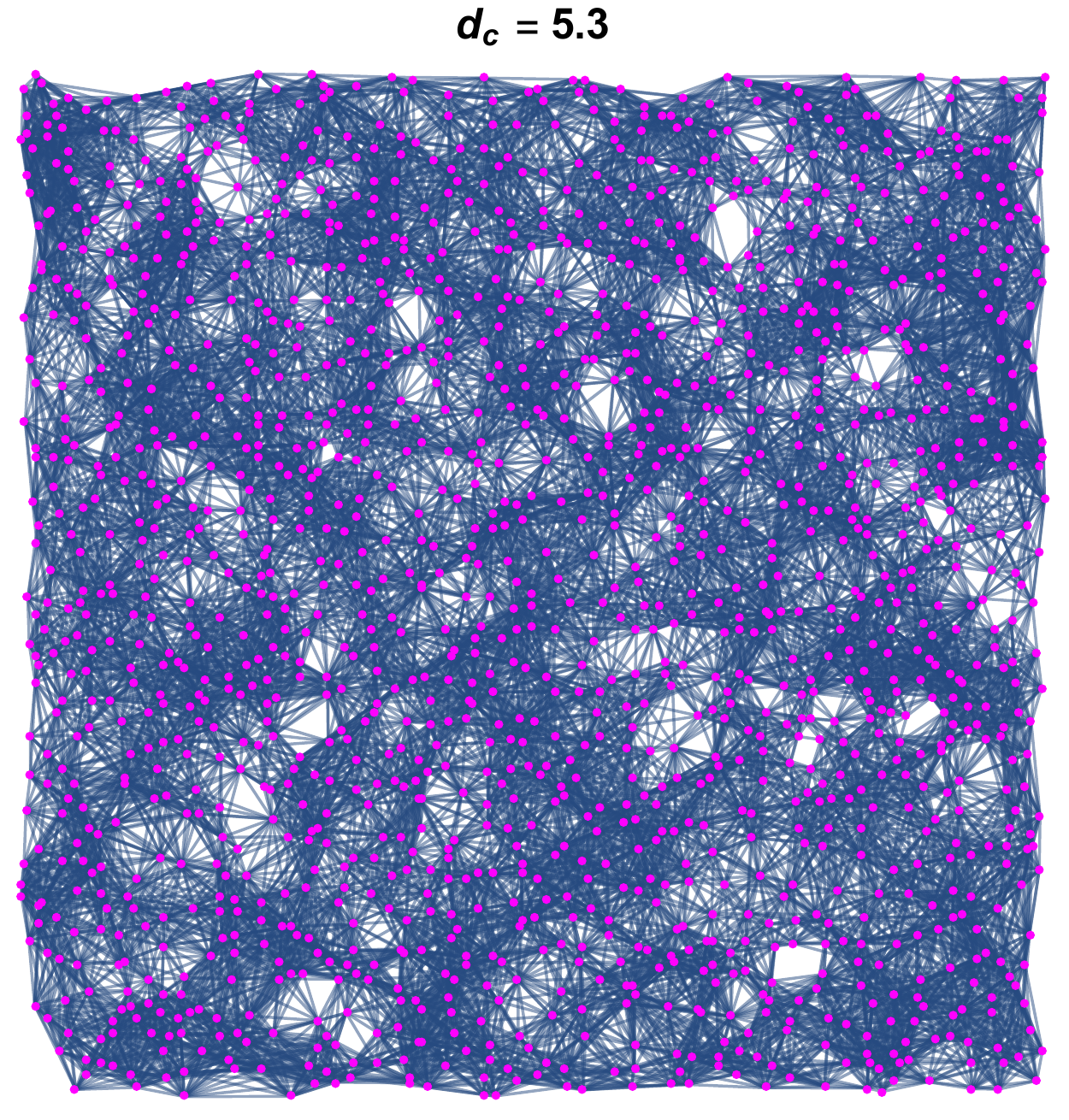}
\caption{\label{fig:graphs} We show the effective graph for one sprinkling with three different choices of $\dcrit$, where the placement of the nodes of the graph (magenta dots) is based on the embedding coordinates. As $\dcrit$ increases, the connectivity across the graph increases. }
\end{figure}

In Fig.~\ref{fig:graphs}, we show the effective graphs $G(\dcrit)$ for various choices of $\dcrit$, to illustrate the appearance of not only small islands, at very small $\dcrit$ but also multiple connectivity of paths at intermediate values of $\dcrit$. However, as $\dcrit$ is increased, this multiple connectivity ``smears'' out and the graph reflects the simple connectedness of an open $d=2$ ball.
 
We extract the spectral dimension from Eq.~\eqref{eq:dscont}, by using a discretised derivative
\be
d_{s,\, n} = - 2 \frac{\sigma_i}{P_r(\sigma_i)} \frac{P_r(\sigma_{i+n})-P_{r}(\sigma_i)}{n}.\label{eq:dsdisc}
\ee
We find that a  choice of  $n=2,3$ smooths out large fluctuations in the return probability (and hence $d_s$) at long diffusion times due to finite statistics.

 \begin{figure}[!b]
\includegraphics[width=0.45\linewidth]{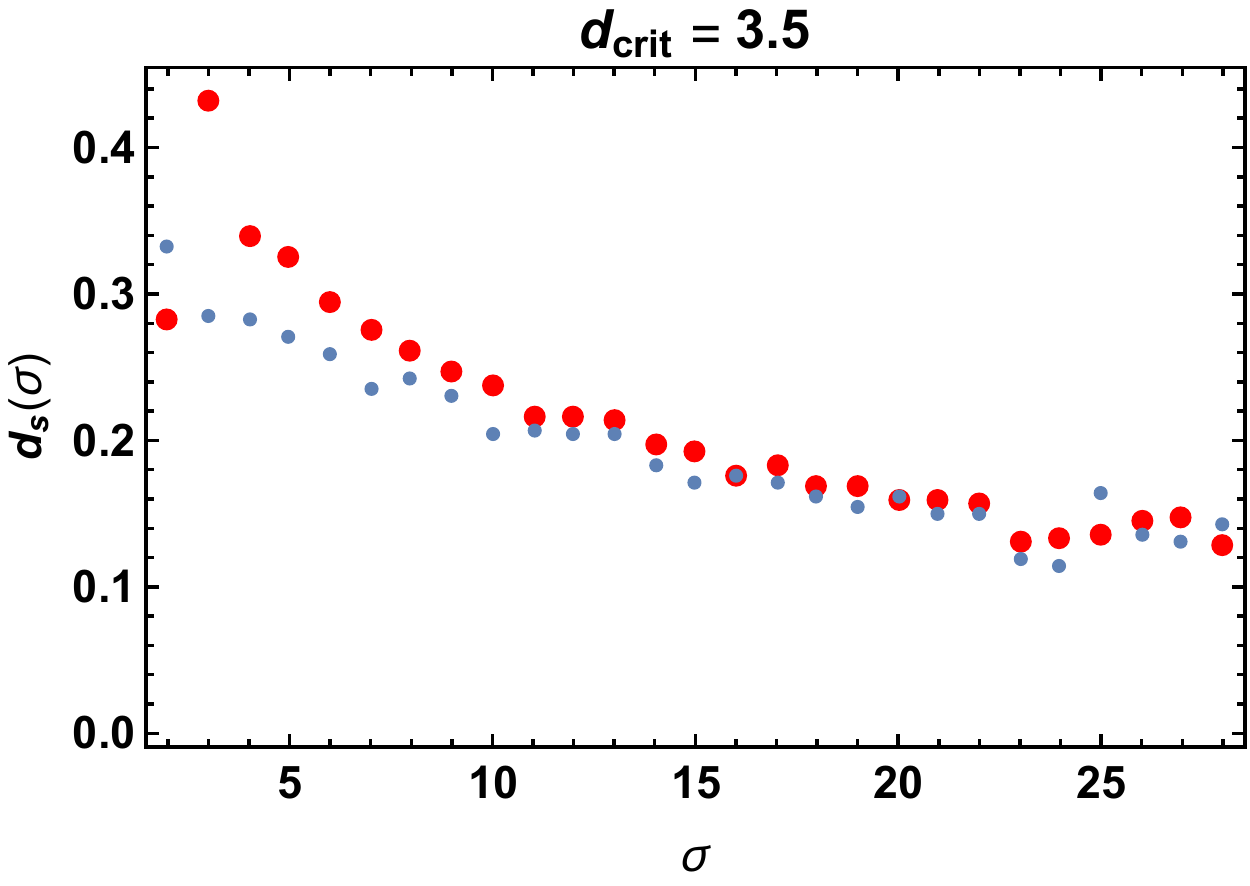} \quad \includegraphics[width=0.45\linewidth]{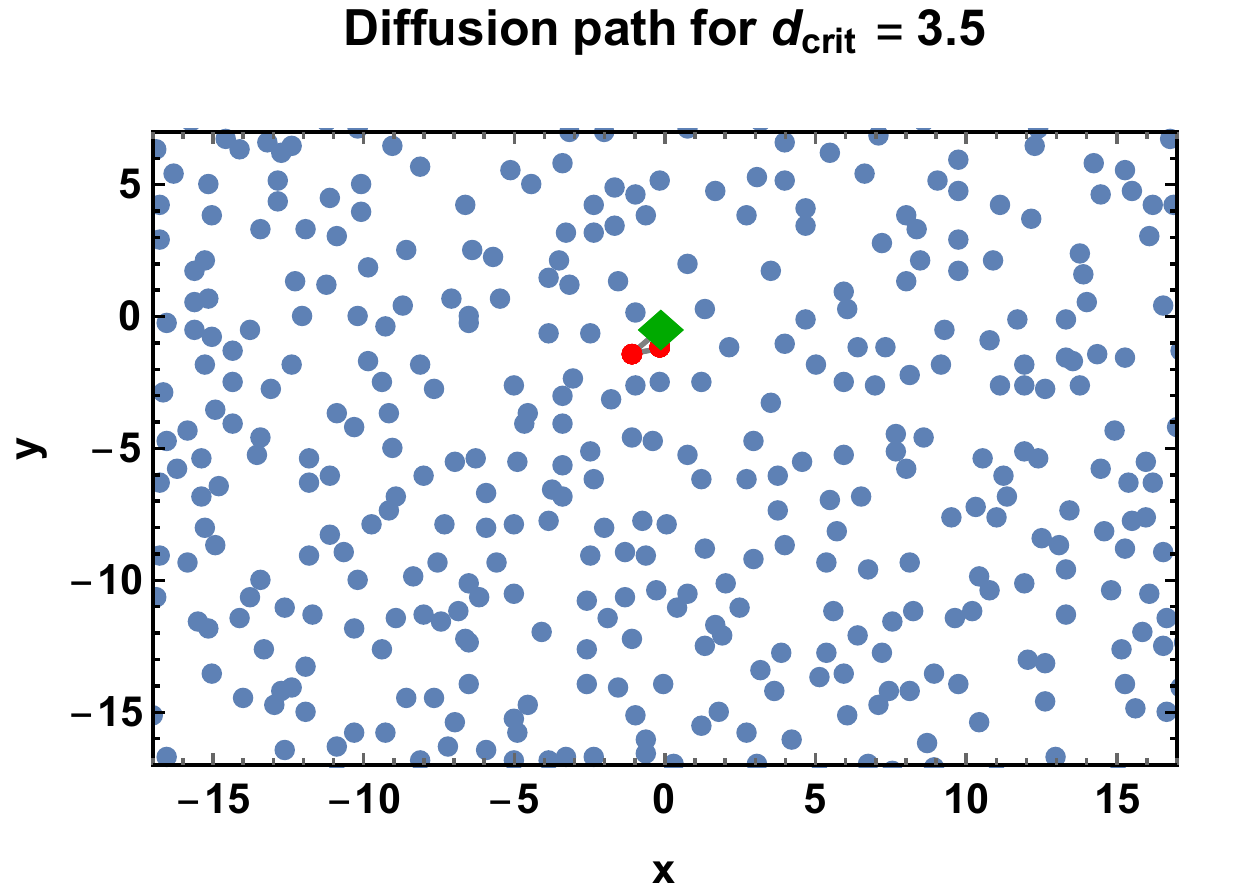}
\caption{\label{fig:ds_35}We show the spectral dimension, extracted from the discretised derivative as in Eq.~\eqref{eq:dsdisc} for $n=2$ (smaller blue dots) and $n=3$ (larger red dots) for $\dcrit = 3.5$ (left panel), and the corresponding diffusion process for a selected sprinkling and starting point (right panel). In the right panel, the blue dots are the elements in the antichain, the green diamond is the starting point and the red dots are the only other two elements that the diffusion process can reach at $\dcrit=3.5$.}
\end{figure}

In Fig.~\ref{fig:ds_35}, we show the spectral dimension for  $\dcrit=3.5$ for the two choices, $n=2,3$. It starts off at approximately $d_s\approx 0.3$ for
very small  $\sigma$ and drops rapidly to smaller values by $\sigma \sim 20$  without showing any indication of a plateau at finite $\sigma$. This is the type of behaviour that is expected once boundary effects become important: $d_s\rightarrow0$ at finite $\sigma$ is a signature of equilibration of the diffusion process. Yet, given the much larger extent of the antichain, one would not expect boundary effects to set in at such small values of  $\sigma$. This puzzle can be resolved by inspecting the diffusion process itself, cf.~right panel in Fig.~\ref{fig:ds_35}. At $\dcrit=3.5$, the antichain consists of isolated ``islands'' with nearest-neighbour-relations on the islands, but none between the separate islands. This traps  the diffusion processes  on small subsets of the antichain. The  diffusion process quickly equilibrates on these subsets, resulting in a continuous drop of the spectral dimension to zero. The existence of the islands and hence the drop-off in $d_s$  is a direct consequence of  asymptotic silence. 

\begin{figure}[!t]
\includegraphics[width=0.45\linewidth]{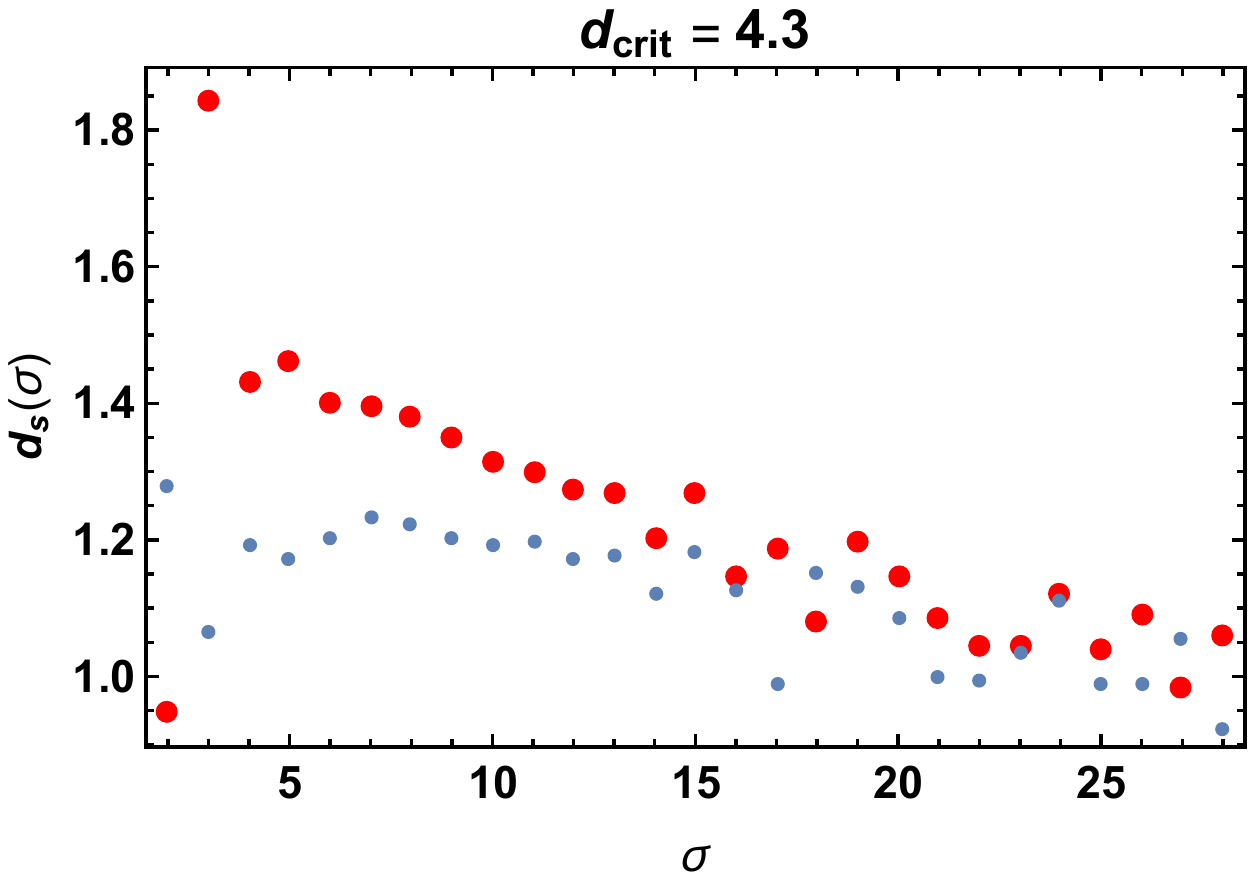} \quad \includegraphics[width=0.45\linewidth]{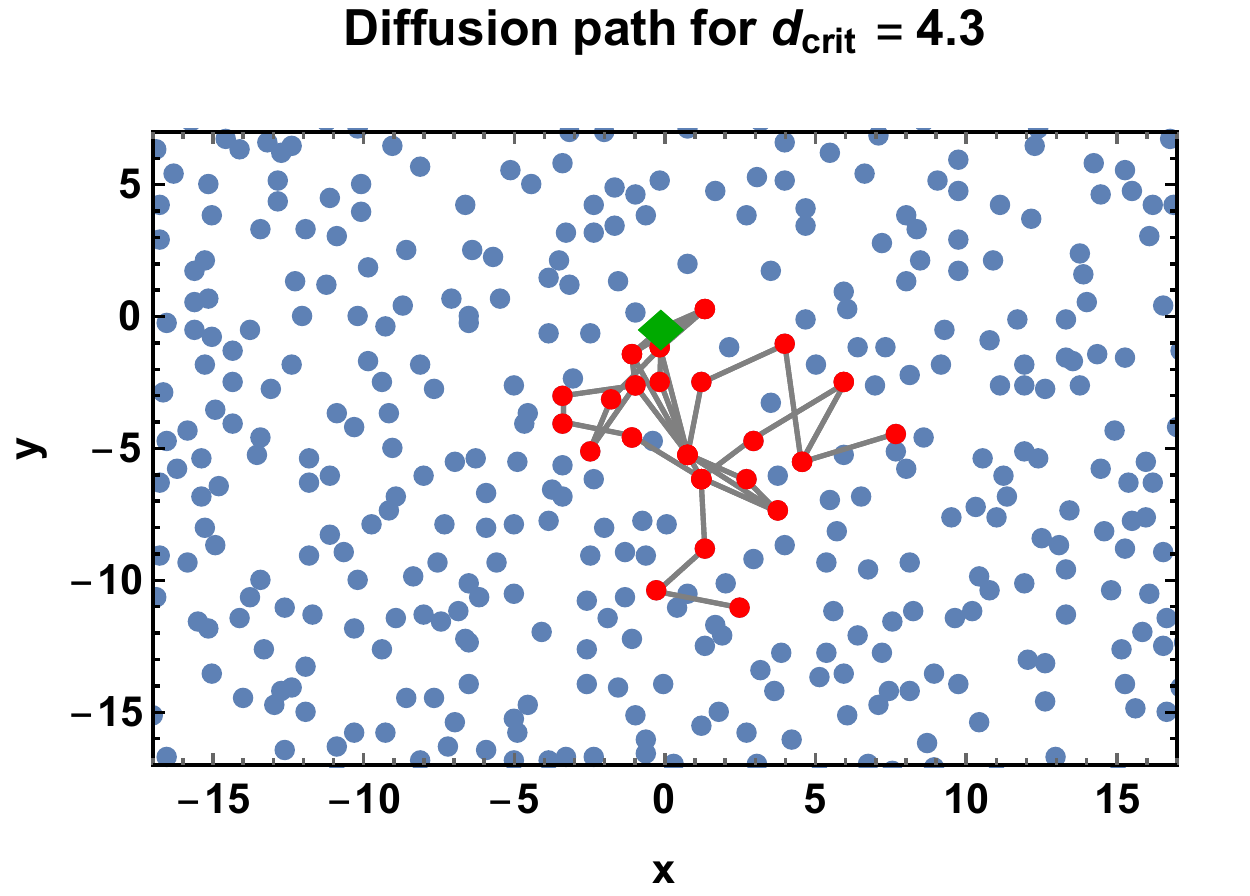}
\caption{\label{fig:ds_43}We show the spectral dimension, extracted from the discretised derivative as in Eq.~\eqref{eq:dsdisc} for $n=2$ (smaller blue dots) and $n=3$ (larger red dots) for $\dcrit = 4.3$ (left panel), and the corresponding diffusion process for a selected sprinkling and starting point (right panel). In the right panel, the blue dots are the elements in the antichain, the green diamond is the starting point and the red dots are the other elements that the diffusion process can reach at $\dcrit=4.3$. One specific diffusion path is indicated.}
\end{figure}

As we increase $\dcrit$, the size  of the islands grows. The increased connectivity within the antichain results in a larger value of $d_s$ at small $\sigma$, cf.~Fig.~\ref{fig:ds_43}. On the other hand,  the fact that $d_s<d_H$ for sufficiently small $\dcrit$ can be interpreted as a dimensional reduction. 

 The reason for the continuous growth of $d_s$ with $\dcrit$ as opposed to a discontinuous transition,  lies in the discrete, random nature of the underlying causal set. Since for a given value of $\dcrit$ the number of nearest neighbours is not a fixed quantity, any causal set will feature patches that are either dense or sparse in the number of elements. At low $\dcrit$, the antichain is just a completely disconnected set of points. As $\dcrit$ is increased, the disconnected islands start to exhibit an increased connectivity. However this does not happen instantaneously, i.e., at a particular $\dcrit$, due to the random nature of the causal set. 
 Consequently, for intermediate $\dcrit$, even in the limit $\sigma \rightarrow \infty$, a diffusion process will not necessarily be able to reach every element in the antichain. Instead, disconnected local neighbourhoods exist within the antichain to which the diffusion process is confined.

\begin{figure}[!t]
 \centering
 \includegraphics[width=0.45\linewidth]{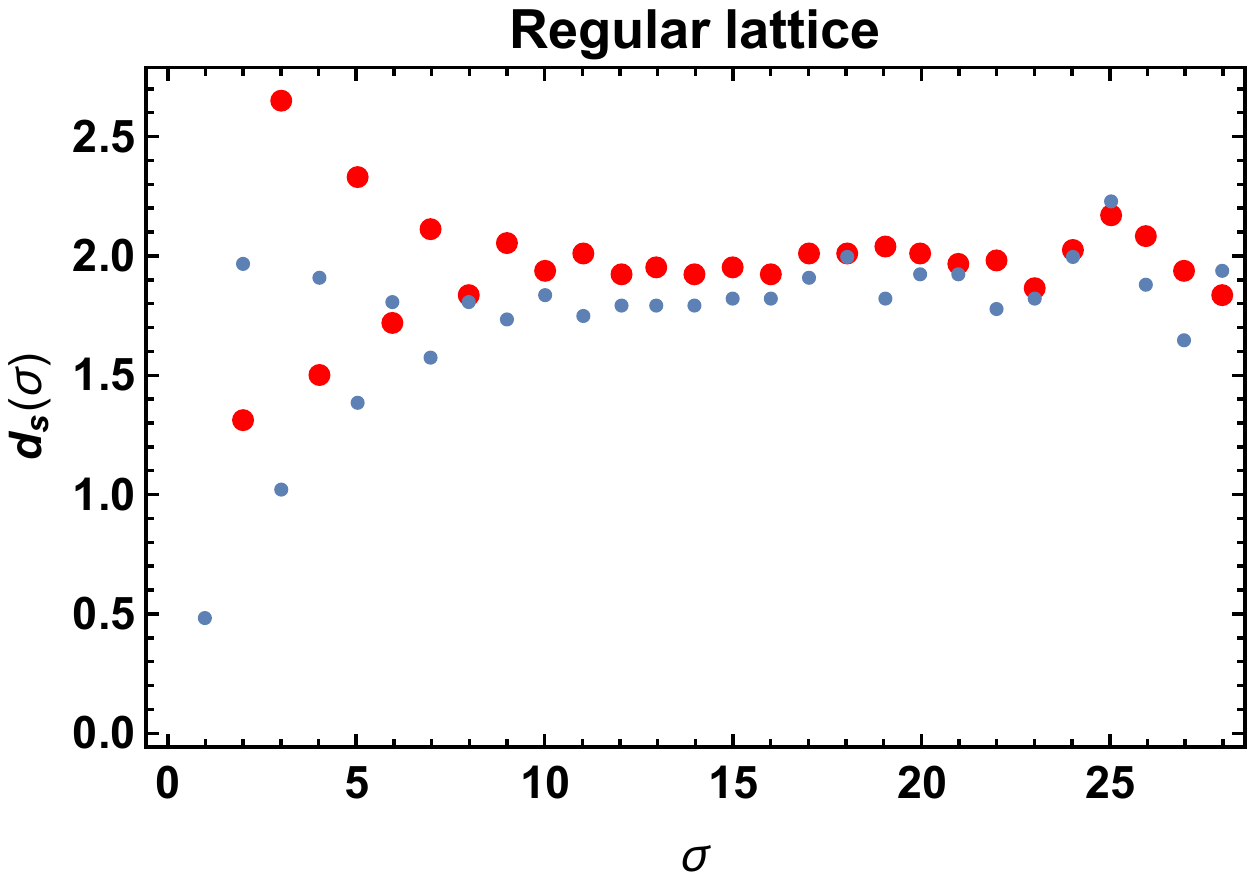}
 \caption{\label{fig:reg_lat} We show the spectral dimension for a regular lattice as a function of $\sigma$. }
 \end{figure}

 We contrast these findings to the case of a finite, regular lattice of lattice-spacing $a$ in two dimensions. For $\dcrit <a$, the lattice consists of isolated islands of only a single element each,  such that the spectral dimension is exactly zero. Once $\dcrit=a$,  global connectivity across the entire regular lattice is achieved, resulting in a spectral dimension that plateaus at $d_s=2$, cf.~Fig.~\ref{fig:reg_lat}.   As a function of $\dcrit$, the spectral dimension at intermediate $\sigma$ therefore sharply jumps from 0 to 2, without exhibiting an intermediate transition regime as in the case of causal sets.
 At small $\sigma$ and $\dcrit \geq a$, one still observes large fluctuations of $d_s$ about the mean $d_s=2$ which is a consequence of the regular nature of the lattice.
 At large $\sigma$, when the diffusion processes reach the boundary, equilibration sets in and results in a drop in the spectral dimension to zero. 
 
\begin{figure}[!t]
\includegraphics[width=0.45\linewidth]{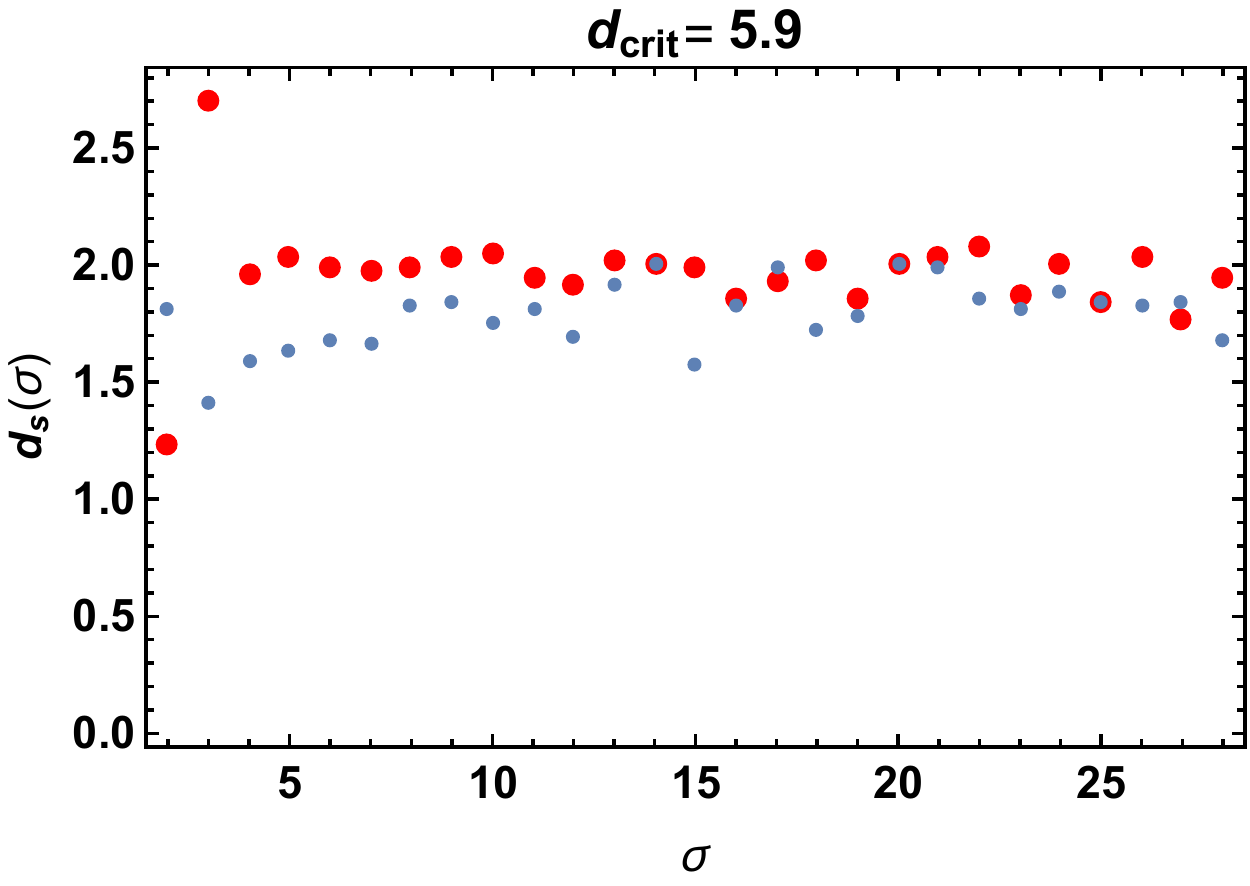}\quad \includegraphics[width=0.4\linewidth]{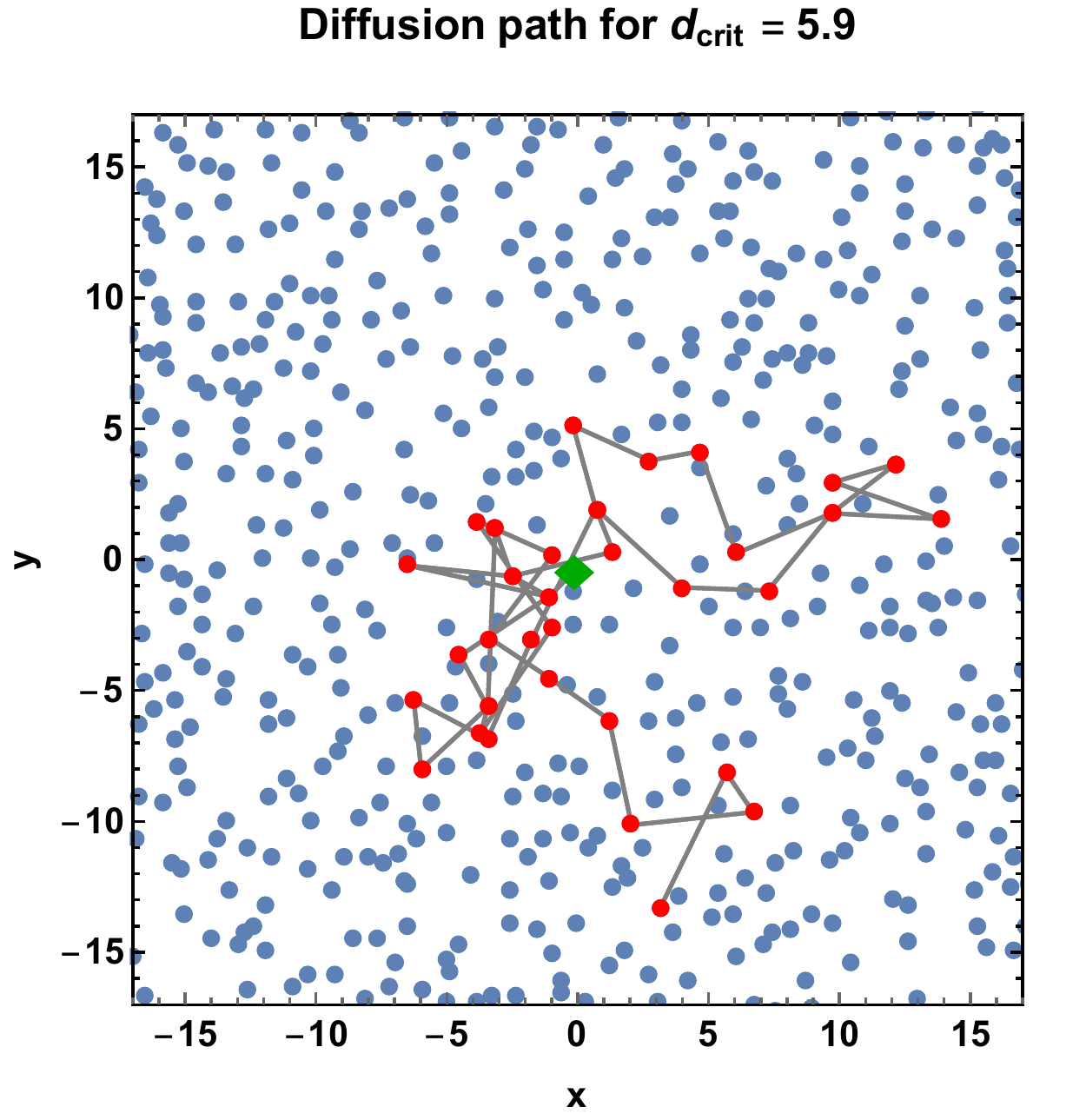}\\
\includegraphics[width=0.45\linewidth]{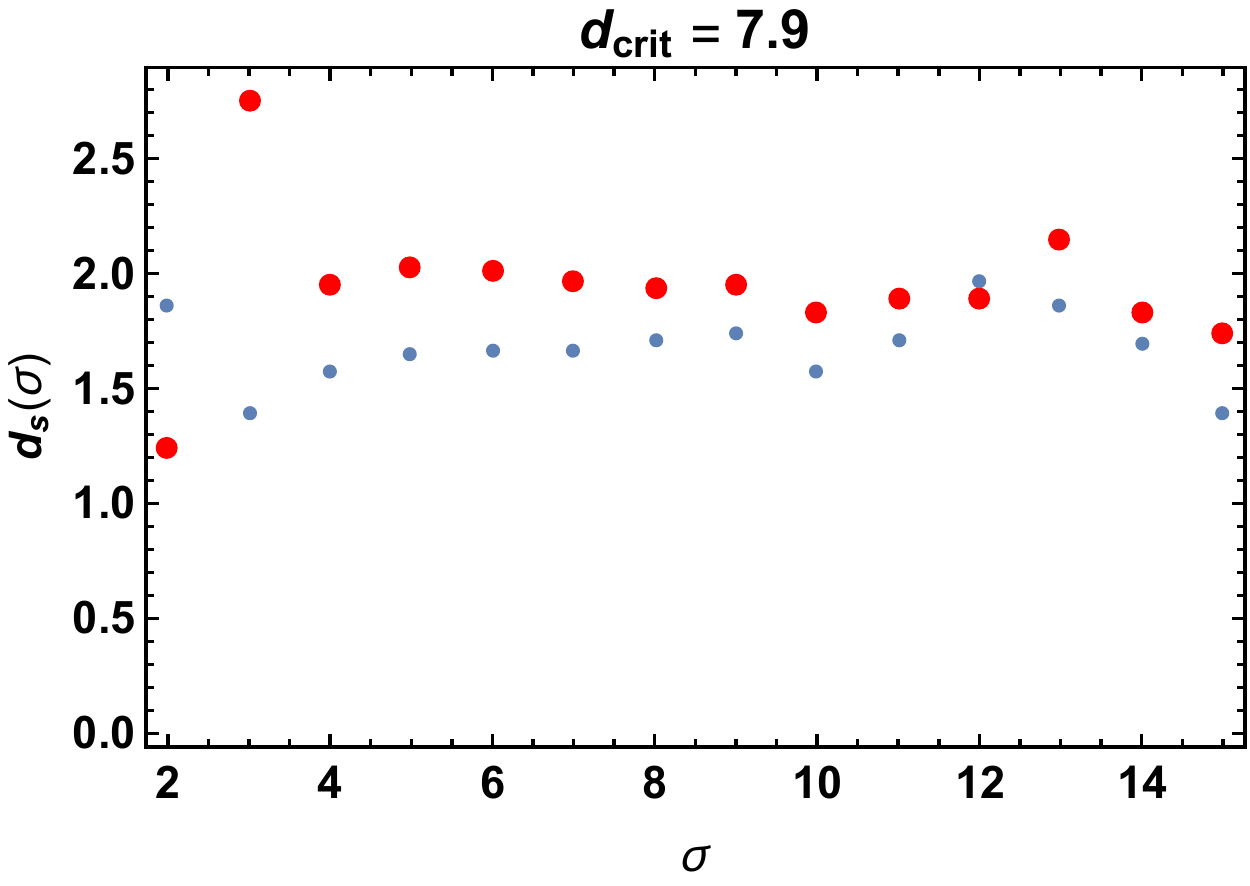}\quad \includegraphics[width=0.4\linewidth]{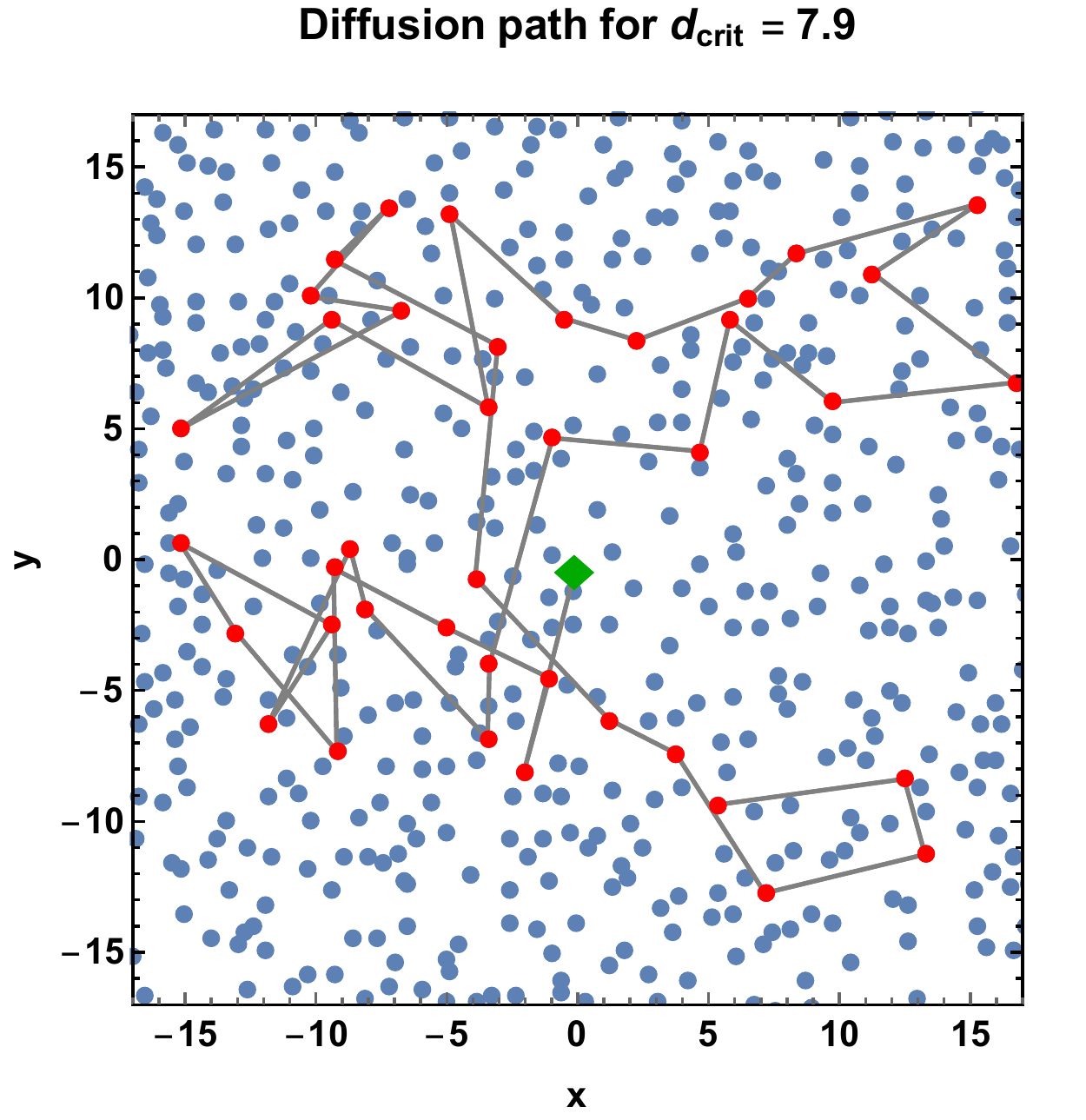}
\caption{\label{fig:d5979} We show the spectral dimension, extracted from the discretised derivative as in Eq.~\eqref{eq:dsdisc} for $n=2$ (smaller blue dots) and $n=3$ (larger red dots) for $\dcrit = 5.9$ (upper left panel) and $\dcrit=7.9$ (lower left panel) as well as corresponding examples of diffusion paths.}
\end{figure}

On the other hand, when  $\dcrit$ is large enough one does find a similar plateauing for the causal set at intermediate $\sigma$, from which the continuum dimension can be read off. We find this to be the case for $\dcrit = 5.9$ and $\dcrit=7.9$, cf.~Fig.~\ref{fig:d5979}. Once connectivity across the complete antichain is achieved, the diffusion process can measure the continuum dimension,  corresponding to the Hausdorff dimension of the antichain. We observe that using $n=3$ vs. $n=2$ in Eq.~\eqref{eq:dsdisc} yields a value closer to $d_s=2$ already  for smaller $\sigma$. This can be attributed to  the fact that $n=3$ is a more coarse-grained derivative, and therefore closer to the continuum approximation. The difference between $n=2$ and $n=3$ is a consequence of the discrete nature of the diffusion process in $\sigma$. 
 Even for larger values of $d_c$, discreteness plays a role at small diffusion times, just as in the regular lattice.

It is important to clearly disentangle the two effects which lead to small $d_s$.  For {\it all} choices of $\dcrit$, small diffusion times tend to show a small spectral dimension, which is a consequence of discreteness. For smaller values of $\dcrit$, the spectral dimension continues to remain small also at intermediate $\sigma$ as  a consequence of discrete asymptotic silence, which confines a diffusion process to a local neighbourhood of its starting point. We interpret this as   dimensional reduction in the UV, since $\dcrit$ determines   the UV or IR  scale.
 
   A further increase in $\dcrit$ leads to the onset of boundary effects at early times $\sigma \approx 6$ so that there is no range of $\sigma$ in which   $d_s$ exhibits a plateau-like behaviour from which  the continuum dimension can be read off. These boundary effects for large $\dcrit$ are simply a consequence of computational limitations. 
 
In summary, $\dcrit$ can be viewed as a measure of scale, such that small values of  $\dcrit$ correspond to diffusion processes which probe the UV regime of the causal set while for larger values of $\dcrit$ they probe the IR regime. We find that while the spectral dimension yields the continuum Hausdorff dimension in the IR, there is a dimensional reduction in the UV.

In \cite{Abajian:2017qub}, using the Myrheim-Meyer dimension estimator \cite{Myrheim:1978ce,Meyer}, it was shown that causal sets that are approximated by  $\mathbb{M}^d$ with  $d=3,4,5$ also show a  dimensional reduction, with the ultraviolet value depending on the dimensional estimator.

Our discussion in this work has been limited to kinematics. However, the dimensional reduction to $d_s=2$ in the UV observed in  several other quantum-gravity approaches, is a consequence of  dynamics, not kinematics. This is perhaps most obvious in the case of asymptotically safe quantum gravity, where $d_s=2$ is a consequence of the UV dynamics becoming scale-invariant. In contrast, studies of causal sets are so far limited to the kinematical regime. Dimensional reduction occurs as a direct consequence of discreteness of single configurations. It is therefore an intriguing question, for which our work has paved the way, whether causal set quantum gravity shows {\it dynamical} dimensional reduction, once the quantum expectation value of the spectral dimension is evaluated. Results in d=2 causal set quantum gravity show the existence of a continuum phase for which the kinematical spectral dimension studied here would continue to be relevant. For the  non-continuum  phase, one would expect drastically modified behaviour \cite{Surya:2011du,Glaser:2014dwa}.  

\begin{acknowledgments}   A.~E.~and F.~V.~are supported by the DFG under the Emmy-Noether program, grant no.~Ei/1037-1. A.~E.~is also supported by the Danish National Research Foundation under grant DNRF:90. S.S. is supported by a Visiting Fellowship (2019-2022) at the Perimeter Institute of Theoretical Physics.
 \end{acknowledgments}

\end{document}